\documentclass[preprintnumbers,aps,prd,twocolumn,superscriptaddress,showpacs]{revtex4-1}

\usepackage{amssymb}
\usepackage{amsmath}
\usepackage{graphicx}
\usepackage{enumerate}
\usepackage{mathtools}
\usepackage{color}

\begin{document}


\preprint{LA-UR-16-25128}

\title{Neutrino flavor transformation in the lepton-asymmetric universe}


\author{Lucas Johns}
\email[]{ljohns@physics.ucsd.edu}
\affiliation{Department of Physics, University of California, San Diego, La Jolla, California 92093, USA}
\affiliation{Theoretical Divison, Los Alamos National Laboratory, Los Alamos, New Mexico 87545, USA}

\author{Mattia Mina}
\affiliation{Theoretical Divison, Los Alamos National Laboratory, Los Alamos, New Mexico 87545, USA}
\affiliation{Department of Physics, University of Torino, Via P. Giuria 1, I - 10125 Torino, Italy}

\author{Vincenzo Cirigliano}
\author{Mark W. Paris}
\affiliation{Theoretical Divison, Los Alamos National Laboratory, Los Alamos, New Mexico 87545, USA}

\author{George M. Fuller}
\affiliation{Department of Physics, University of California, San Diego, La Jolla, California 92093, USA}


\date{\today}

\begin{abstract}
We investigate neutrino flavor transformation in the early universe in the presence of a lepton asymmetry, focusing on a two-flavor system with $1-3$ mixing parameters.  We identify five distinct regimes that emerge in an approximate treatment neglecting collisions as the initial lepton asymmetry at high temperature is varied from values comparable to current constraints on the lepton number down to values at which the neutrino--neutrino forward-scattering potential is negligible.  The characteristic phenomena occurring in these regimes are (1) large synchronized oscillations, (2) minimal flavor transformation, (3) asymmetric ($\nu$- or $\bar{\nu}$-only) MSW, (4) partial MSW, and (5) symmetric MSW.  We examine our numerical results in the framework of adiabaticity, and we illustrate how they are modified by collisional damping.  Finally, we point out the existence of matter--neutrino resonances in the early universe and show that they suffer from non-adiabaticity.
\end{abstract}


\maketitle


\section{Introduction}

In this paper we examine how cosmological lepton asymmetries spawned at high temperature affect the ensuing evolution of neutrino flavor.  Despite the particle's humble stature, the consequences of neutrino physics for the early universe are profound.  As the universe cools to a temperature of a few MeV, the weak-interaction rates that have safeguarded thermal equilibrium in the neutrino sector begin to falter in their competition with Hubble expansion.  At roughly the same time, electrons and positrons are annihilating and dumping entropy into the plasma.  Some neutrinos share in this heating, but not all --- leaving their once-equilibrium spectra deformed and cool compared to those of photons, which remain in equilibrium by dint of their swift electromagnetic interactions.  (See, for example, Ref.~\cite{grohs2016} for a recent discussion of the Boltzmann transport of neutrino energy and entropy through weak decoupling and Big Bang nucleosynthesis (BBN).)

During this period neutrinos are all the while undergoing capture on free nucleons and contributing to blocking factors in electron/positron capture and neutron decay.  Through their role in these processes, neutrinos shape the neutron-to-proton ($n/p$) ratio that will be available when the nucleus-building begins in full force at $T \sim 70$ keV.  The primordial byproducts of BBN --- most promisingly, from an observational perspective, the elements D and $^4$He --- depend on the $n/p$ ratio, and the protracted freeze-out of weak interactions means that there is ample time for the evolving, non-equilibrium neutrino spectra to leave their mark on the nuclide abundances \cite{grohs2016b}.

Even after neutrinos have decoupled from the plasma, they are no mere spectators, as their energy density helps to set the expansion rate of the universe.  In the era following $e^{\pm}$ annihilation, neutrinos are relativistic and therefore contribute, along with photons (and possibly other particles beyond the Standard Model), to the radiation energy density $\rho _\textrm{rad}$.  The energy density of these species is commonly parameterized in terms of the quantity $N_{\textrm{eff}}$, defined by the relation
\begin{equation}
\rho _\textrm{rad} = 2 \left[ 1 + \frac{7}{8} \left( \frac{4}{11} \right) ^{4/3} N_\textrm{eff} \right] \frac{\pi ^2}{30} T^4.
\end{equation}
This parameter is sensitive not just to the number of flavors of neutrinos but also to their post-decoupling spectra, which, as noted above, inevitably sustain non-thermal distortions.  Determining the precise form of these distortions and their impact on BBN and $N_\textrm{eff}$ is a rich and persistent problem in cosmology \cite{dodelson1992, dolgov1992, dolgov1997, *dolgov1999, gnedin1998, lopez1999, esposito2000, mangano2002, mangano2005, grohs2016, grohs2016b}.

Of particular importance in this regard is the lepton number
\begin{equation}
L_\nu = \frac{n_\nu - n_{\bar{\nu}} }{n_\gamma},
\end{equation}
defined in terms of the number densities of neutrinos ($n_\nu$), antineutrinos ($n_{\bar{\nu}}$), and photons ($n_\gamma$).  In thermal equilibrium a finite lepton number is tantamount to one or more nonzero chemical potentials in the neutrino sector, with clear ramifications for $N_\textrm{eff}$.  Away from equilibrium the chemical potentials are no longer well-defined, but the implications of nonzero $L_\nu$ for the radiation energy density still stand.  A cosmological lepton number also exerts an influence through the special role, indicated previously, that the electron flavor plays in mediating the reactions
\begin{align}
\nu_e + n & \rightleftharpoons p + e^- \notag \\
\bar{\nu}_e + p & \rightleftharpoons n + e^+.
\end{align}
The unique leverage on the primordial $^4$He abundance that $\nu_e$ and $\bar{\nu}_e$ are afforded by virtue of these reactions \cite{kneller2004} has driven interest in the possibility that $L_\nu$ is not only nonzero but is (or once was) distributed unevenly across the individual flavors.  The evolution of an initial lepton asymmetry --- a difference between $L_{\nu_e}$ and $L_{\nu_x}$ in the effective two-flavor scenario that we will investigate --- depends on the interplay between collisions and medium-enhanced oscillations, both of which are capable of shuttling lepton number between flavors.  In a lepton-asymmetric universe especially, precise predictions of $N_\textrm{eff}$ and $Y_P$ (the mass fraction of $^4$He) therefore demand a careful treatment of neutrino flavor transformation.

Due to the influence of sphalerons, the lepton number is expected in many baryogenesis models to be comparable to the baryon asymmetry (or baryon-to-photon ratio) $\eta = n_B / n_\gamma \approx 6 \times 10^{-10}$ \cite{kuzmin1985, arnold1987, *arnold1988, harvey1990}.  But the fact remains that the lepton number is only modestly constrained by measurements: Even the most stringent bounds currently permit asymmetries of $\sim 5 \times 10^{-2}$ \cite{dolgov2002, serpico2005, simha2008, mangano2012, castorina2012, steigman2012}, a full eight orders of magnitude above $\eta$.  Moreover, the past several decades have brought forth a number of models \cite{harvey1981, dolgov1991, foot1996, shi1996, casas1999, march1999, mcdonald2000, kawasaki2002, yamaguchi2003, shaposhnikov2008, gu2010} that can generate a large lepton number without contravening the impressive agreement on $\eta$ between cosmic microwave background (CMB) and BBN data.  A measurement of the lepton number of the universe, whatever its value turns out to be, will serve as a probe of physics at and above the scale of electroweak symmetry breaking and will put to the test theories of baryogenesis.

As of recently, a careful treatment is now motivated from yet another direction. The detections \cite{bulbul2014, boyarsky2014} of a mysterious X-ray line in a number of galaxies and galaxy clusters at $\sim 3.55$ keV have ignited speculation that the line may be attributable to dark matter decay.  One scenario consistent with this interpretation --- indeed, a scenario that may be said to have predicted the appearance of a keV decay line \cite{abazajian2001} --- is the resonant production of sterile neutrino dark matter in the presence of a nonzero lepton number \cite{shi1999, abazajian2001b, kishimoto2008}.  (For reviews of the dark-matter candidacy of sterile neutrinos, see Refs.~\cite{kusenko2009, adhikari2016}.)  Given the energy and flux of the alleged decay line, resonant production singles out a range of pre-resonance lepton numbers on the order of $L_\nu \sim 5 \times 10^{-4}$ as being in best agreement with the X-ray observations \cite{abazajian2014}.  Since the production mechanism is agnostic to the details of how $L_\nu$ is distributed, it leaves the door open to lepton asymmetries and any signatures that they may have left behind.

Investigation into the evolution of the individual lepton numbers dates back at least to the work of Savage, Malaney, and Fuller \cite{savage1991}, who considered the role of resonant neutrino oscillations --- a topic that will be a major theme of the present work.  But the current orthodoxy on the subject originated a decade later with the watershed numerical study by Dolgov et al. \cite{dolgov2002} and the papers by Abazajian et al. \cite{abazajian2002} and Wong \cite{wong2002} that followed shortly thereafter.  (See also Ref.~\cite{lunardini2001}.)  The authors of Ref.~\cite{dolgov2002} concluded that equilibration of the lepton number across the flavors --- the shorthand for which is simply \textit{flavor equilibration} --- is achieved prior to the onset of BBN for a lepton asymmetry on the order of the $L_\nu$ constraint.  Subsequent papers on the topic \cite{abazajian2002, wong2002, pastor2009, gava2010, mangano2011, mangano2012, castorina2012} have refined this original treatment of the problem, examining the connections to $N_\textrm{eff}$, $Y_P$, and the D abundance $\left[ \textrm{D} / \textrm{H} \right]$.

The literature in this area has largely been inspired by the quest to establish rigorous limits on the neutrino degeneracy parameters $\eta_{\nu_\alpha} = \mu_\alpha / T$, where $\mu_\alpha$ is the chemical potential of neutrino flavor $\alpha$.  In the event that a lepton number completely equilibrates, the BBN-derived limits that constrain $\eta_{\nu_e}$ likewise apply to the other flavors.  Conversely, if no equilibration occurs, then the constraints on $\eta_{\nu_\mu}$ and $\eta_{\nu_\tau}$ are considerably weaker than those on $\eta_{\nu_e}$, as they are bounded solely by their contribution to the radiation energy density.  The objective of this paper is not to revisit the question of constraints on neutrino degeneracy, but rather to explore more fully the panoply of flavor evolution that may have occurred in the early universe.  While smaller values of $L_\nu$ push into the realm of effects that are thought to be currently undetectable, we demonstrate --- with an eye to forthcoming observational improvements --- that varying the initial lepton number leads to dramatically different behaviors.

\begin{figure}
\includegraphics[width=.44\textwidth]{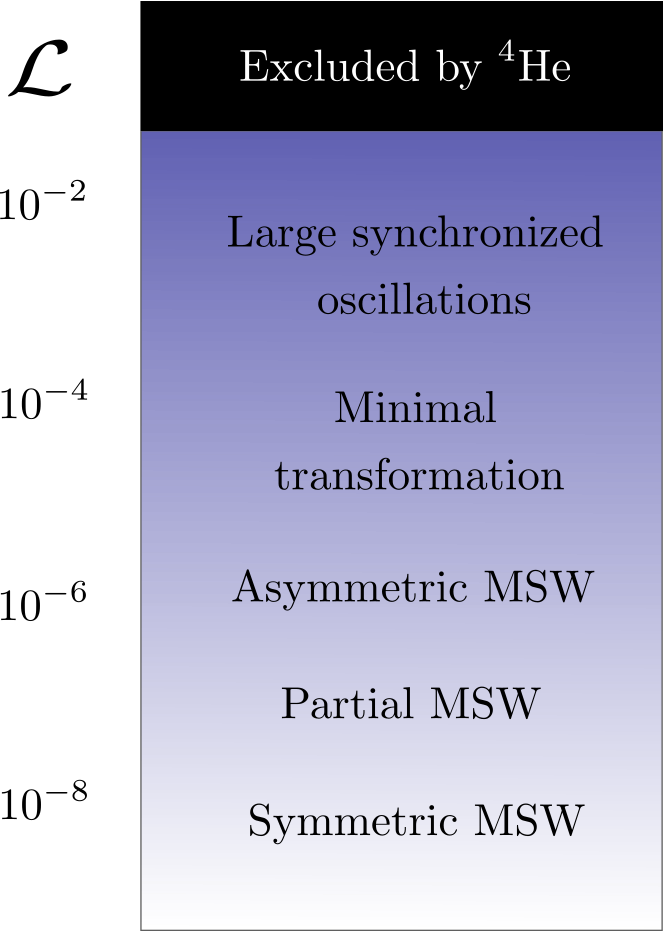}
\caption{(Color online) Schematic illustration of the landscape of coherent flavor evolution in the inverted hierarchy as a function of lepton asymmetry $\mathcal{L} = L_{\nu_e} - L_{\nu_x}$.  The black swath at the top of the figure indicates the realm of lepton asymmetries that are currently excluded by $^4$He measurements.  The five regimes at sub-constraint values of $\mathcal{L}$ are labeled by their most prominent characteristics.  \label{regimefig}}
\end{figure}

To this end we identify five regimes of coherent flavor evolution that may be found for lepton numbers at or below observational constraints (Fig.~\ref{regimefig}).  We label these regimes by their principal characteristics, which are, in order of decreasing lepton asymmetry, 
\begin{enumerate}[I.]
	\item large synchronized oscillations,
	\item minimal flavor transformation,
	\item asymmetric ($\nu$- or $\bar{\nu}$-only) MSW conversion,
	\item partial MSW conversion, and
	\item symmetric MSW conversion.
\end{enumerate}
We elucidate the physics behind these behaviors with frequent recourse to the framework of (non-)adiabatic level-crossings.  We also discuss how the inclusion of collisions, in the approximate form of scattering-induced quantum damping, differently affects these regimes.  Coarse features of the coherent oscillation physics are found to persist in the presence of damping.  We argue that this finding motivates further exploration of the lepton-asymmetric terrain with a treatment that goes beyond the approximations of the present study.

The equations of motion relevant to the evolution of a lepton asymmetry in the early universe are set out in Sec.~\ref{kinetics}.  The regimes of coherent evolution are presented in Sec.~\ref{results}, followed by discussions of adiabaticity, the matter--neutrino resonance, and the importance of collisions.  A conclusion is given in Sec.~\ref{conclusion}.  Throughout this paper we use natural units in which $c = \hbar = k_B = 1$.

\section{Neutrino kinetics in the early universe \label{kinetics}}

For reasons made clear below, the active period for neutrino flavor transformation begins around $10-20$ MeV and continues down to --- or, depending on the lepton asymmetry, through --- the epoch of neutrino decoupling at $\sim 1$ MeV.  Over these temperatures the three flavors of active neutrinos are immersed in a hot, dense bath of electrons, positrons, and free nucleons; the $\mu^{\pm}$ and $\tau^{\pm}$ that abounded at higher temperatures have all but disappeared, while $e^{\pm}$ remain relativistic through to the very bottom of this temperature range.  Protons and neutrons, in contrast, have long since become nonrelativistic, and their densities are minuscule in comparison on account of the high entropy of the plasma.  Within this medium neutrinos experience oscillations enhanced by forward (coherent) scattering with matter particles ($e^{\pm}$) and other neutrinos.  They also undergo momentum-changing (incoherent) scattering with both populations.

In Sec.~\ref{2flavor} we explain how the problem of neutrino flavor evolution under these conditions can be reduced to an effective two-flavor scenario.  We then go on to describe in Sec.~\ref{subkinetics} the potentials that drive the coherent mixing between the two flavors and the incoherent scattering that competes against it.  We provide in Sec.~\ref{subresonant} the relevant background on resonant flavor transformation and collective oscillations.  Lastly, in Sec.~\ref{numissues} we briefly summarize the numerical approach adopted in this study.

\subsection{Two-flavor system \label{2flavor}}

Under the condition that $L_{\nu_\mu} = L_{\nu_\tau}$, the paucity of muons and tauons in the plasma --- and, correspondingly, of charged-current interactions involving $\nu_\mu$ and $\nu_\tau$--- entails that neutrinos are well modeled by an effective two-flavor system consisting of $\nu_e$ and $\nu_x$, where $\nu_x$ is a superposition of $\nu_\mu$ and $\nu_\tau$.  We present here a derivation of the effective mixing parameters relevant to this two-flavor system.  A similar view on the reduction to two flavors can be found in Ref.~\cite{balantekin1999}.

An effective mixing angle $\theta$ parameterizes vacuum mixing in the two-flavor system, with the orthogonal transformation between the mass and flavor states given by
\begin{align}  &\nu_e = \nu_1 \cos\theta + \nu_2 \sin\theta \notag \\
&\nu_x = -\nu_1 \sin\theta + \nu_2 \cos\theta , \end{align}
where $\nu_{1,2}$ are mass eigenstates with masses $m_{1,2}$.  (Strictly speaking, as Eq.~\ref{nu2nu1} below will make clear, $\nu_1$ is only an eigenstate in the limit that two of the three physical neutrino masses are degenerate.  This is exactly the limit that we will take.)  The electron neutrino $\nu_e$ in the two-flavor system is identical to its three-flavor counterpart $\nu'_e$, which after transforming to the appropriate mass bases yields the constraint
\begin{align} 
\nu_1 \cos\theta + \nu_2 \sin\theta & =  \nu'_1 \cos\theta'_{12} \cos\theta'_{13} \notag \\  
& ~~~ + \nu'_2 \sin\theta'_{12}\cos\theta'_{13} e^{-i \varphi_1} \notag \\
& ~~~ + \nu'_3 \sin\theta'_{13} e^{-i \delta} e^{-i \varphi_2}, 
\end{align}
using primes to denote three-flavor mixing parameters and $\delta$ and $\varphi_i$ to denote the Dirac and Majorana CP-violating phases.  We identify $\theta = \theta'_{13}$, so that the two-flavor mass eigenstates are related to the three-flavor ones by the relations
\begin{align} &\nu_1 = \nu'_1 \cos\theta'_{12} + \nu'_2 \sin\theta'_{12} e^{-i \varphi_1} \notag \\
&\nu_2  = \nu'_3 e^{-i\delta} e^{-i \varphi_2}. \label{nu2nu1} \end{align}
The phases $\delta$ and $\varphi_2$ amount to an overall rephasing of $\nu_2$ and exert no influence on our calculations; similarly for the other Majorana phase.  We point out that this conclusion regarding $\delta$ is consistent with the study of CP violation in the neutrino-degenerate early universe in Ref.~\cite{gava2010}, which showed that effects of CP violation from the Dirac phase appear only when $L_{\nu_\mu} \neq L_{\nu_\tau}$.

A third mass eigenstate, orthogonal to $\nu_1$ and $\nu_2$ and having mass $m_3$, may also be defined in order to complete the transformation between the primed and unprimed bases:
\begin{equation}
\nu_3 = -\nu '_1 \sin\theta ' _{12} + \nu '_2 \cos\theta ' _{12} e^{-i \varphi_1}. \label{nu3}
\end{equation}
This state decouples from the other two and is identical to the third flavor eigenstate in the unprimed basis: $\nu_3 = \nu_y$.  Written in terms of the physical flavor states,
\begin{align}
& \nu_e = \nu'_e \notag \\
& \nu_x = \nu'_\mu \sin\theta'_{23} e^{-i\delta} + \nu'_\tau \cos\theta'_{23} e^{-i\delta} \notag \\
& \nu_y = \nu'_\mu \cos\theta'_{23} - \nu'_\tau \sin\theta'_{23}.
\end{align}
From Eqs.~\eqref{nu2nu1} and \eqref{nu3} it follows that 
\begin{align}
&m^2_1 = m'^2_1 \cos^2\theta'_{12} + m'^2_2 \sin^2\theta'_{12} \notag \\
&m^2_2 = m'^2_3 \notag \\
&m^2_3 = m'^2_1 \sin^2\theta'_{12} + m'^2_2 \cos^2\theta'_{12}.
\end{align}
Since we are concerned primarily with qualitative behavior in this paper, we will be content to take $m'^2_1 \approx m'^2_2$, which leads to
\begin{equation}
\delta m^2 \equiv m_2^2 - m_1^2 \approx m_2^2 - m_3^2 \approx \delta m'^2_{31}.
\end{equation}
To this level of approximation the $\nu_i$ ($i = 1, 2, 3$) states are genuine mass eigenstates and, moreover, $\nu_3$ is degenerate with $\nu_1$ and decouples from the $\nu_e - \nu_x$ mixing channel.  With the proviso that the lepton numbers $L_{\nu_\mu}$ and $L_{\nu_\tau}$ are the same (but not necessarily equal to $L_{\nu_e}$), the flavor transformation that occurs in the temperature range we investigate here is therefore adequately captured by $\nu_e - \nu_x$ oscillations with $1-3$ mixing parameters.

This effective two-flavor system distills many of the important aspects of the full three-flavor problem, and the flavor-transformation phenomena we describe below carry over to mixing in other channels.  The locations and sizes of features shift with changing parameters --- $\delta m_\odot ^2$, for instance, gives rise to resonance behavior at lower temperatures than does $\delta m_\textrm{atm} ^2$ --- but the physics behind these features is resilient.  Nonetheless, it should be kept in mind that transformation among three flavors will lead to an even richer landscape of flavor evolution than in the two-flavor scenario, especially in the event that $L_{\nu_e}$, $L_{\nu_\mu}$, and $L_{\nu_\tau}$ are all unequal.

\subsection{The kinetic equations \label{subkinetics}}

Tracking the flavor content of an ensemble of neutrinos and antineutrinos is accomplished by following the evolution of the density matrices $\rho$ and $\bar{\rho}$, which for each comoving energy $\epsilon = E / T \approx p / T$ have the $2 \times 2$ structures
\begin{equation}
\rho \left( \epsilon, t \right) = \left( \begin{array}{cc}
\rho_{ee} & \rho_{ex} \\
\rho^*_{ex} & \rho_{xx}
\end{array} \right), ~ \bar{\rho} \left( \epsilon, t \right) = \left( \begin{array}{cc}
\bar{\rho}_{ee} & \bar{\rho}_{ex} \\
\bar{\rho}^*_{ex} & \bar{\rho}_{xx}
\end{array} \right),
\end{equation}
where the individual matrix elements tacitly depend on $\epsilon$ and $t$.  (Throughout this paper we denote the analogous objects for antineutrinos using the prescription $\nu_\alpha \rightarrow \bar{\nu}_\alpha$.  The antineutrino analogues will always be denoted with an overbar.)

We choose a normalization such that at high temperature $\rho$ assumes the form
\begin{equation}
\rho \left( \epsilon \right) ~~ \cong ~~ \left( \begin{array}{cc}
f \left( \epsilon, \eta_{\nu_e} \right) & 0 \\
0 & f \left( \epsilon, \eta_{\nu_x} \right) \label{rhoeq}
\end{array} \right),
\end{equation}
where the diagonal entries are Fermi--Dirac equilibrium distribution functions
\begin{equation}
f \left( \epsilon, \eta_{\nu_\alpha} \right) = \frac{1}{e^{\epsilon - \eta_{\nu_\alpha}}  + 1}. \label{fdeq}
\end{equation}
In general, whether at high temperature or not, the diagonal entries of $\rho$ encode the number densities of $\nu_e$ and $\nu_x$ and the off-diagonal entries measure quantum coherence between the two flavors.

The initial conditions given in Eq.~\eqref{rhoeq} are justified by the quasi-equilibrium that obtains at the starting temperatures used for our calculations.  At these temperatures the neutrinos exchange energy with the plasma on timescales short compared to the Hubble time, ensuring that the neutrino spectra retain their thermal Fermi--Dirac shape on the latter timescale, even while the chemical potentials are evolving.  (To be precise, it is the number densities in \textit{energy} eigenstates that are proportional to Fermi--Dirac functions; the validity of using them in the \textit{flavor}-basis density matrix lies in the fact that at high $T$ the flavor and energy bases are nearly coincident.)  As the temperature drops, oscillations grow in importance relative to incoherent scattering, and the ability of scattering to preserve equilibrium spectra diminishes.  But in the scattering-dominated limit, in which our initial temperatures safely fall, neutrinos have distribution functions as in Eq.~\eqref{rhoeq}, and coherence between the flavors is efficiently stamped out by the high scattering rate.

For each mode $\epsilon$ the neutrino and antineutrino density matrices obey the equations of motion
\begin{align}
&i \left( \partial_t - H p \hspace{1 pt} \partial_p \right) \rho \left( \epsilon, t \right) = \left[ \mathcal{H} \left( \epsilon, t \right), \rho \left( \epsilon, t \right) \right] + \mathcal{C} \notag \\
&i \left( \partial_t - H p \hspace{1 pt} \partial_p \right) \bar{\rho} \left( \epsilon, t \right) = \left[ \bar{\mathcal{H}} \left( \epsilon, t \right), \bar{\rho} \left( \epsilon, t \right) \right] + \bar{\mathcal{C}}, \label{eom}
\end{align}
where $H$ is the Hubble parameter, $\mathcal{H}$ is the Hamiltonian, and $\mathcal{C}$ is the collision term encapsulating incoherent scattering \cite{sigl1993}.  The collision term depends on the neutrino density matrices and the background-particle distribution functions across all energies.

The Hamiltonian consists of three ingredients: a vacuum potential $\mathcal{H}_\textrm{vac}$, which is driven by the mass-squared splitting $\delta m^2$ and the vacuum mixing angle $\theta$; a thermal potential $\mathcal{H}_e$, which is due to forward scattering of neutrinos with the $e^{\pm}$ jostling about in the plasma; and a self-coupling potential $\mathcal{H}_\nu$, which arises from neutrino--neutrino scattering.  Written out,
\begin{align}
\mathcal{H} & = \mathcal{H}_\textrm{vac} + \mathcal{H}_e + \mathcal{H}_\nu \notag \\
& = \frac{\delta m^2}{2 E} \mathrm{B} - \frac{8 \sqrt{2} G_F E \varrho_{e^{\pm}}}{3 m_W^2} \mathrm{L} \notag \\ 
& ~~~ + \frac{\sqrt{2} G_F}{2 \pi^2} \int dE' E'^2 \left[ \rho \left( E' \right) - \bar{\rho}^* \left( E' \right) \right], \label{ham}
\end{align}
where in the flavor basis $\mathrm{B} = \mathrm{U} \left( \textrm{diag} \left[ -1/2, 1/2 \right] \right) \mathrm{U}^\dagger$ with Pontecorvo--Maki--Nakagawa--Sakata (PMNS) matrix $\mathrm{U}$, $\mathrm{L} = \textrm{diag} \left[ 1, 0 \right]$, $\varrho_{e^{\pm}}$ denotes the energy density of $e^{\pm}$, $G_F$ is the Fermi constant, and $m_W$ is the $W$ boson mass.  The time-dependence of $E$, $\varrho_{e^{\pm}}$, $\rho$, and $\bar{\rho}$ is implicit.  Antineutrinos, meanwhile, evolve under the Hamiltonian $\bar{\mathcal{H}} = \mathcal{H}_\textrm{vac} + \mathcal{H}_e - \mathcal{H}_\nu^*$.

Strictly speaking, the Hamiltonian relevant to neutrino propagation in a medium contains more terms than those shown in Eq.~\eqref{ham} \cite{notzold1988}.  In addition to the finite-temperature charged-lepton potential ($\sim \varrho_{e^\pm}$) and the finite-density neutrino potential ($\sim \left( n_{\nu_e} - n_{\nu_x} \right)$ for the diagonal portion), neutrinos also experience a finite-temperature neutrino potential ($\sim \left( \varrho_{\nu_e} - \varrho_{\nu_x} \right)$, again for the diagonal portion) and a finite-density charged-lepton potential ($\sim \left( n_{e^-} - n_{e^+} \right)$).  By the charge neutrality of the universe, however, the $e^\pm$ asymmetry must balance the baryon asymmetry, making this contribution to the potential very small.  The thermal neutrino potential, meanwhile, is $\mathcal{O}(G_F^2)$ and is further suppressed by a factor comparable to the lepton asymmetry.  Lastly, we leave out the $\mu^\pm$ contribution to $\mathcal{H}_e$, as in the scenarios we are concerned with, their population has dwindled close to zero by the time flavor transformation begins.

The collision term $\mathcal{C}$ in Eq.~\eqref{eom} represents inelastic scattering of neutrinos and is proportional to $G_F^2$.  A fully realistic treatment would involve computing quantum Boltzmann collision integrals \cite{vlasenko2014, blaschke2016}, a task that has only recently been accomplished for the first time \cite{desalas2016}.  Whereas in Ref.~\cite{desalas2016} de Salas and Pastor executed a high-precision calculation of $N_\textrm{eff}$ in the standard (i.e., lepton-symmetric) scenario, our aim here is to point out that an initial lepton asymmetry at high temperature shapes the subsequent neutrino flavor evolution in diverse and complicated ways.  For this study we instead set $\mathcal{C}$ to be a quantum damping term that is proportional to $\rho$ but has vanishing diagonal entries \cite{harris1981, *harris1982, stodolsky1987, thomson1992, raffelt1993, mckellar1994, bell1999}.  Using such a term for $\mathcal{C}$ amounts to the ansatz that the chief effect of collisions is to eliminate coherence between the flavors.

The paradigm typically associated with quantum damping holds that a collision acts as a measurement of the scattered neutrino, thereby collapsing it into a definite flavor state.  Although this picture is only a heuristic and has its limitations, it correctly suggests that a system of (anti)neutrinos immersed in a thermal bath ultimately approaches a mixed state with equal $\nu_e$ ($\bar{\nu}_e$) and $\nu_x$ ($\bar{\nu}_x$) probabilities.  One of the fundamental issues at stake with a lepton asymmetry is the timescale over which this descent to a maximum-entropy state (and the concomitant flavor equilibration) transpires.  Conceptual aid notwithstanding, damping does not in fact capture all of the microphysics of scattering, and in Sec.~\ref{dampingsec} we will address the deficiencies of this approximation at length.  

Rather than solving for the flavor evolution directly as a function of $t$, we work in terms of a parameter $x = Ma$, where $M$ is an arbitrary energy scale and $a$ is the scale factor; doing so transfigures the equations of motion into ordinary differential equations.  Furthermore, for a two-flavor system the density matrix $\rho$ can be projected onto the Pauli matrices according to
\begin{equation}
\rho = \frac{1}{2} \left(P_0 + \vec{P} \cdot \vec{\sigma} \right).
\end{equation}
Given that $\mathcal{C}$ has vanishing diagonal entries, the trace of $\rho$ is preserved by the equations of motion and has value
\begin{equation}
\textrm{Tr} \rho = P_0 = f \left( \eta^i_{\nu_e} \right) + f \left( \eta^i_{\nu_x} \right), \label{tracerho}
\end{equation}
where $f \left( \eta^i_{\nu_\alpha} \right)$ is the initial distribution function of $\nu_\alpha$, prior to any significant flavor transformation.  The polarization vector $\vec{P}$, meanwhile, does evolve: If similar projections are performed for $\mathcal{H}$ and $\mathcal{C}$, Eq.~\eqref{eom} can be recast as
\begin{equation}
Hx \frac{d \vec{P}}{dx} = \vec{\mathcal{H}} \times \vec{P} - \mathcal{D} \vec{P}_T. \label{veceom}
\end{equation}
Along the same lines, we will make use of the notation
\begin{equation}
\mathcal{H} = \left( \begin{array}{cc}
\mathcal{H}_z & \mathcal{H}_T \\
\mathcal{H}^*_T & -\mathcal{H}_z
\end{array} \right),
~
\mathcal{V} = \left( \begin{array}{cc}
\mathcal{V}_z & \mathcal{V}_T \\
\mathcal{V}^*_T & -\mathcal{V}_z
\end{array} \right), \label{hvmats}
\end{equation}
where $\mathcal{V} = \mathcal{H}_e + \mathcal{H}_\nu$ denotes the weak-interaction potential arising from coherent forward scattering and where, for example, $\mathcal{H}_T = \mathcal{H}_x - i \mathcal{H}_y$ encodes the component of the Hamiltonian vector $\vec{\mathcal{H}}$ that is transverse to the flavor ($z$-) axis.  The damping parameter $\mathcal{D}$ that appears in Eq.~\eqref{veceom} is related to the scattering amplitudes of the neutrino flavor states.  To illustrate: If the medium were such that the two flavors had equal scattering amplitudes, interactions would be unable to differentiate between the two flavors and there would be no damping ($\mathcal{D} = 0$).  At the other extreme, if one of the flavors were non-interacting (for instance, in active--sterile mixing), then the damping parameter would be half the total interaction rate $\Gamma_\alpha$ of the other flavor ($\mathcal{D} = \Gamma_\alpha / 2$).

In the plasma of the early universe, $\nu_e$ and $\nu_x$ scatter with different (but nonzero) cross sections, and a detailed derivation of $\mathcal{D}$ would add up the contributions from all the individual weak-interaction processes relevant to this environment.  We opt for a coarser treatment here, taking 
\begin{equation}
\mathcal{D} \approx \frac{1}{2} \left( \Gamma_e - \Gamma_x \right) \approx \frac{1}{2} d_{ex} G_F^2 p T^4 \label{dampapprox}
\end{equation}
with $d_{ex} \approx 0.35$ \cite{bell1999}.  The approximation in Eq.~\eqref{dampapprox} is sufficiently accurate for the objectives of this study.  Indeed, for most of the paper we focus on the coherent regime, in which $\mathcal{D} = 0$; it is only in Sec.~\ref{dampingsec} that we let $\mathcal{D}$ assume the approximate form given above.  To be sure, the flavor evolution we are tracking occurs over a range of temperatures in which collisions are important.  But as we discuss in detail in Sec.~\ref{dampingsec}, broad characteristics of the coherent regime survive the inclusion of damping, and a close analysis of the coherent transformation sheds light on the physics underlying the behavior in the presence of both oscillations and collisions.

Translating the density matrix $\rho$ into the polarization vector $\vec{P}$ affords a geometric interpretation to the flavor evolution of the system (more detailed expositions of which may be found in Refs. \cite{stodolsky1987, kim1988}).  At high temperatures $\vec{P}$ lies along the $z$-axis because of the peremptory destruction of coherence by collisions.  As the temperature (and by extension the scattering rate) drops, $\vec{P}$ is able to travel away from the $z$-axis: The path it follows is determined by a competition between its desire to precess around the Hamiltonian vector $\vec{H}$ and the constant push exerted by collisions back toward the flavor axis.  Meanwhile $\vec{H}$ itself migrates in response to the falling temperature and the movement of the individual polarization vectors.  $\vec{P}$ tries to track $\vec{H}$ as the latter drifts, but its success in doing so is moderated by the constant buffeting of collisions and the degree of non-adiabaticity.  We address the latter criterion in Sec.~\ref{subadiabaticity}.

At any time the relative number density of $\nu_e$ and $\nu_x$ of a given mode $\epsilon$ (taken to have finite width $d\epsilon$) can be read off by projecting that mode's polarization vector onto the flavor axis and providing the appropriate thermodynamic prefactor:
\begin{align}
& P_{z, \epsilon} = \rho_{ee, \epsilon} - \rho_{xx, \epsilon} \notag \\
\Longrightarrow  ~ & dn_{\nu_e, \epsilon} - dn_{\nu_x, \epsilon}  = \frac{T^3}{2 \pi^2} ~d\epsilon ~\epsilon ^2 P_{z,\epsilon}. \label{pzdef}
\end{align}
We have written the number density of $\nu_\alpha$ in mode $\epsilon$ as $dn_{\nu_\alpha, \epsilon}$ in preparation for integrating over all modes.  Performing the sum over $\epsilon$ and dividing by $T^3$ (to get a redshift-invariant quantity) yields the $z$-component of the \textit{integrated} polarization vector:
\begin{equation}
P_{z,\textrm{int}} \equiv \frac{1}{2 \pi^2} \sum_\epsilon d\epsilon ~\epsilon ^2 P_{z,\epsilon} = \frac{n_{\nu_e} - n_{\nu_x}}{T^3},
\end{equation}
where $n_{\nu_e}$ and $n_{\nu_x}$ are the total number densities across all energies.  The quantities $\bar{P}_z$ and $\bar{P}_{z, \textrm{int}}$, appropriate to antineutrinos, are defined analogously.  We will present many of our numerical results as plots of $P_{z,\textrm{int}}$ and $\bar{P}_{z, \textrm{int}}$, as they convey the ``average'' flavor evolution of the system; where illuminating, we will zoom in on the individual modes.  Note that with these definitions $P_z > 0$ ($\bar{P}_z > 0$) reflects a predominance of electron neutrinos (antineutrinos).

\subsection{Resonant flavor mixing and collective oscillations \label{subresonant}}

One of the linchpins of neutrino flavor phenomenology in the early universe and other astrophysical environments is the Mikheyev--Smirnov--Wolfenstein (MSW) mechanism by which coherent scattering with a matter background causes neutrinos to acquire effective masses and mixing angles \cite{wolfenstein1978, mikheyev1985}.  Letting $\Delta \equiv \delta m^2 / 2 E$, the in-medium mass-squared splitting $\delta m^2 _M$ is defined by
\begin{equation}
\Delta ^2_M \equiv \left( \frac{\delta m^2 _M}{2 E} \right)^2 \equiv  \Delta^2 \sin^2 2\theta + \left( \Delta \cos 2\theta - \mathcal{V}_z \right) ^2 \label{D2M}
\end{equation}
and the in-medium mixing angle $\theta_M$ by
\begin{equation}
\sin^2 2\theta_M \equiv \frac{\Delta^2 \sin^2 2\theta}{\Delta^2 _M}. \label{s2M}
\end{equation}
(For the purposes of introducing the traditional MSW mechanism, we are neglecting neutrino--neutrino scattering in Eqs.~\eqref{D2M} and \eqref{s2M}, but we will return to these definitions later on in order to incorporate self-coupling.)  Resonance occurs when $\mathcal{V}_z = \Delta \cos 2 \theta$: The effective mixing angle is at its maximum (to wit, $\theta_M = \pi/4$) and the effective mass-squared splitting at its minimum.  Since the thermal potential $\mathcal{H}_e$ (Eq.~\eqref{ham}) depends only on the energy density of $e^{\pm}$ in the plasma, the matter background modifies the oscillations of neutrinos and antineutrinos in precisely the same way.

Neutrino--neutrino coherent scattering gives rise to ``index-of-refraction'' effects in much the same fashion as a matter background, but with an added layer of complexity.  As seen in Eq.~\eqref{ham}, the evolution of $\rho(\epsilon)$ for a particular mode $\epsilon$ depends, through the self-coupling potential $\mathcal{H}_\nu$, on the density matrices for all other modes $\epsilon '$, meaning that the problem of flavor evolution in a dense neutrino system is a nonlinear one.  A fascinating range of collective behaviors has been shown to result.  (See Ref.~\cite{duan2010} for a review, or Refs.~\cite{cherry2012, raffelt2013, mirizzi2013, vlasenko2014b, keister2015, abbar2015} for a selection of recent work in this active area.)  The role of nonlinear coupling in the early universe is perhaps best epitomized by the phenomenon of synchronized oscillations that emerges when the self-coupling is strong enough to ``glue'' all of the individual modes together and prevent them from kinematically decohering \cite{samuel1993, kostelecky1993, kostelecky1993b, *kostelecky1994, *kostelecky1995, pastor2002}.

Synchronized oscillations are seen in our results to be one of the hallmarks of coherent flavor evolution in a universe with a lepton asymmetry within a couple orders of magnitude of the current constraint on $L_\nu$.  At the other end of the spectrum, with a lepton asymmetry on the order of the baryon asymmetry $\eta$, self-coupling is unimportant and the MSW mechanism reigns supreme.  In the following section we discuss these two regimes and several others that emerge at intermediate lepton asymmetries.

\begin{figure}
\includegraphics[width=.47\textwidth]{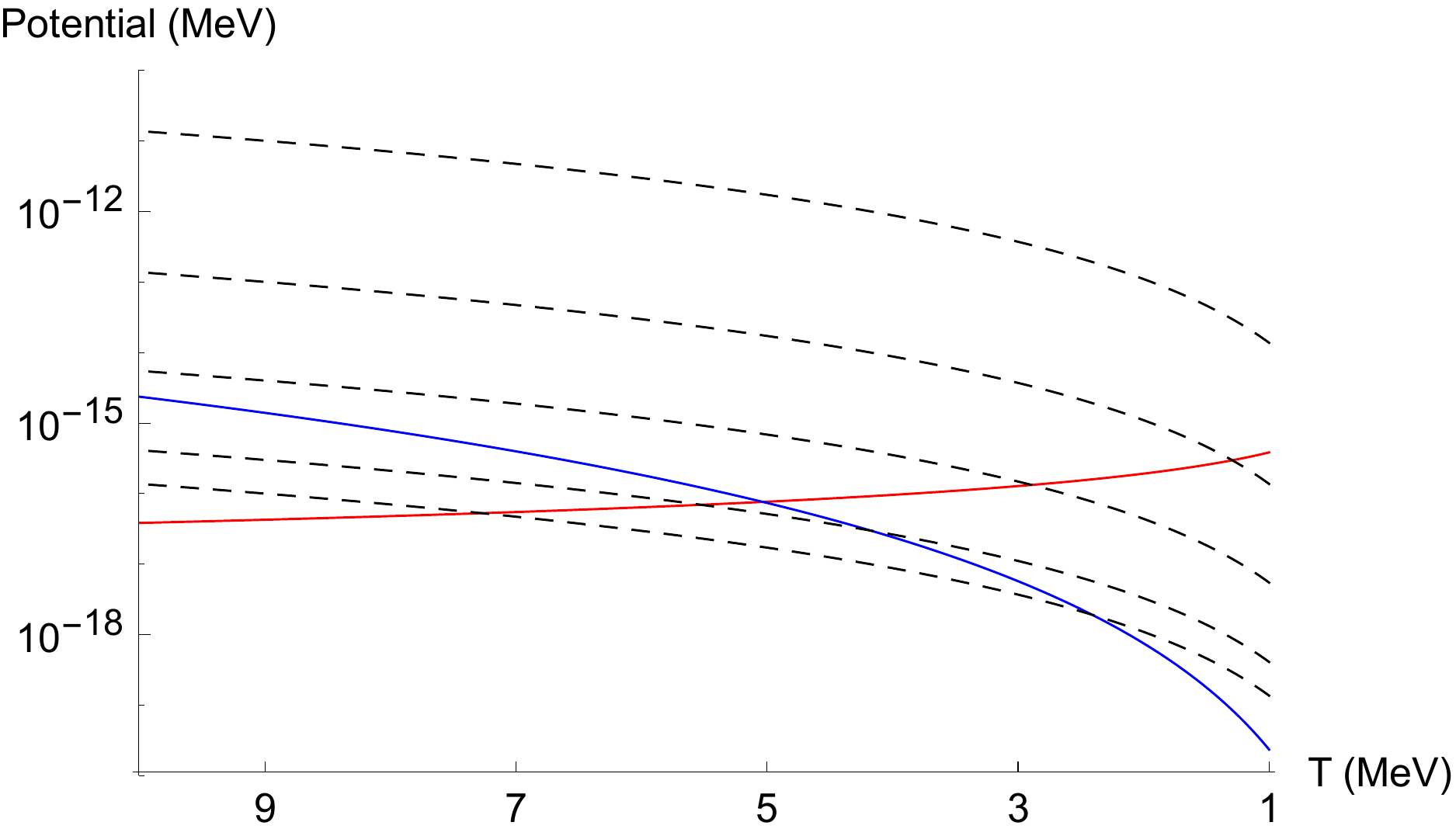}
\caption{(Color online) Magnitudes of the individual diagonal potentials for $\epsilon = 3$: $| \mathcal{H}_\textrm{vac, z} |$ (red, upper solid curve at $T = 1$ MeV), $| \mathcal{H}_{e,z} |$ (blue, other solid curve), and $|\mathcal{H}_{\nu, z} |$ (black, dashed).  The last of these was computed assuming no flavor transformation.  From top to bottom the dashed curves correspond to degeneracy parameters $\eta_{\nu_e} = 5 \times 10^{-3}$, $5 \times 10^{-5}$, $2 \times 10^{-6}$, $1.5 \times 10^{-7}$, and $5 \times 10^{-8}$ (Eq.~\eqref{defdeg}); chemical potentials in $\nu_\mu$ were taken to be zero. \label{potentials}}
\end{figure}

The type of behavior exhibited depends fundamentally on the relative sizes of the individual contributions to the Hamiltonian.  We depict in Fig.~\ref{potentials} the magnitudes of the diagonal potentials as functions of temperature, with four different curves for $\mathcal{H}_{\nu, z}$ corresponding to different initial lepton asymmetries.  As we describe below, one of the basic determinants of the flavor evolution is the magnitude of $\mathcal{H}_{\nu, z}$ where the vacuum- and thermal-potential curves intersect, which is to say at MSW resonance.  We will also see the limitations of this picture, which fails to account for the off-diagonal components of the Hamiltonian.

\subsection{Numerical details \label{numissues}}

We have employed two independent programs for solving the equations of motion (Eq.~\eqref{veceom}), one based on a fourth-order explicit Runge--Kutta solver and the other on a Magnus-method solver.  The Magnus method, which is tailor-made for tracking unitary evolution, has been used previously in work on neutrino flavor transformation in supernovae; for a thorough description, see Ref.~\cite{duan2008}.  We have achieved consistent results with the two codes, and we have confirmed that in the coherent limit each one individually conserves $| \vec{P} |$ and $\textrm{Tr} \rho = P_0$ to high precision.

As shown below, certain flavor-evolution regimes host rapid, highly aperiodic oscillations, and in such regimes the behavior of individual modes depends sensitively on the physical and computational parameters of the calculation.  The very fine features displayed in these scenarios are without (and may simply defy) a detailed physical explanation and, moreover, are beyond the level of precision aimed at in this study.  Rather, our focus is on the major qualitative features, which we have found to be robust.

\section{Results and Discussion \label{results}}

In this section we present our results through the example of five different initial lepton asymmetries that typify the major regimes of coherent flavor evolution in the inverted hierarchy (IH).  We then apply the concept of adiabaticity to gain insight into the behaviors manifested in these prototypical cases.  Following a discussion of the coherent regimes, we introduce collisions in the form of quantum damping.  As a rule of thumb, the impact of damping is (in a non-quantitative sense) proportional to the amount of flavor transformation that would occur in the \textit{absence} of damping: That is to say, for damping to gain leverage on the evolution of $\vec{P}$, a significant $\vec{P}_T$ must develop, and for this to be the case there must be substantial transformation of $\vec{P}$ away from the initial flavor eigenstate.  To understand the results with damping, it is therefore necessary to understand the results without.

In what follows we focus most of our attention on the IH because it, unlike the normal hierarchy (NH), plays host to an MSW resonance and, by implication, to generally more substantial flavor transformation.  We will briefly discuss the NH when we turn to damping.

\subsection{Regimes of coherent evolution}

Before any flavor transformation has occurred neutrinos of flavor $\alpha$ are described by a Fermi--Dirac equilibrium spectrum with neutrino degeneracy parameter $\eta_{\nu_\alpha}$ (Eq.~\eqref{fdeq}).  For the purposes of this study we assume that at high temperatures the lepton number is positive and entirely contained in $\nu_e$, so that $\eta_{\nu_x} = 0$ and $\eta_{\nu_e}$ can be deduced from the lepton number via
\begin{equation}
L_\nu \approx \frac{1}{12 \zeta \left( 3 \right) } \left( \pi^2 \eta_{\nu_e} + \eta_{\nu_e}^3 \right) \approx 0.68 \eta_{\nu_e}, \label{defdeg}
\end{equation}
where the second approximation applies for the small degeneracy parameters we are considering.  We would find similar results, but with the roles of neutrinos and antineutrinos swapped, if instead we were to take a negative $\eta_{\nu_e}$ or were to put the lepton number entirely in $\nu_x$.  Furthermore, the choice of setting $\eta_{\nu_x} = 0$ at high temperature is inessential for our results, as it is the lepton $\textit{asymmetry}$ which dictates the role of the self-coupling potential.

Note that we take no stance on what mechanism actually produces the initial lepton numbers.  The question of how to generate an asymmetry that survives washout from scattering processes is an important one and has been examined in Ref.~\cite{ghiglieri2016}.  This question is, however, outside the purview of the present study.

The five regimes of coherent flavor evolution that we have identified in our numerical results are depicted schematically in Fig.~\ref{regimefig}.  In our sweep of the lepton-number terrain, the values of $\eta_{\nu_e}$ that we have found best embody the features associated with these regimes are as follows: $5 \times 10^{-8}$, $1.5 \times 10^{-7}$, $2 \times 10^{-6}$, $5 \times 10^{-5}$, and $5 \times 10^{-3}$.

\subsubsection{$\eta_{\nu_e} = 5 \times 10^{-8}$: Symmetric MSW}

\begin{figure}
\includegraphics[width=.47\textwidth]{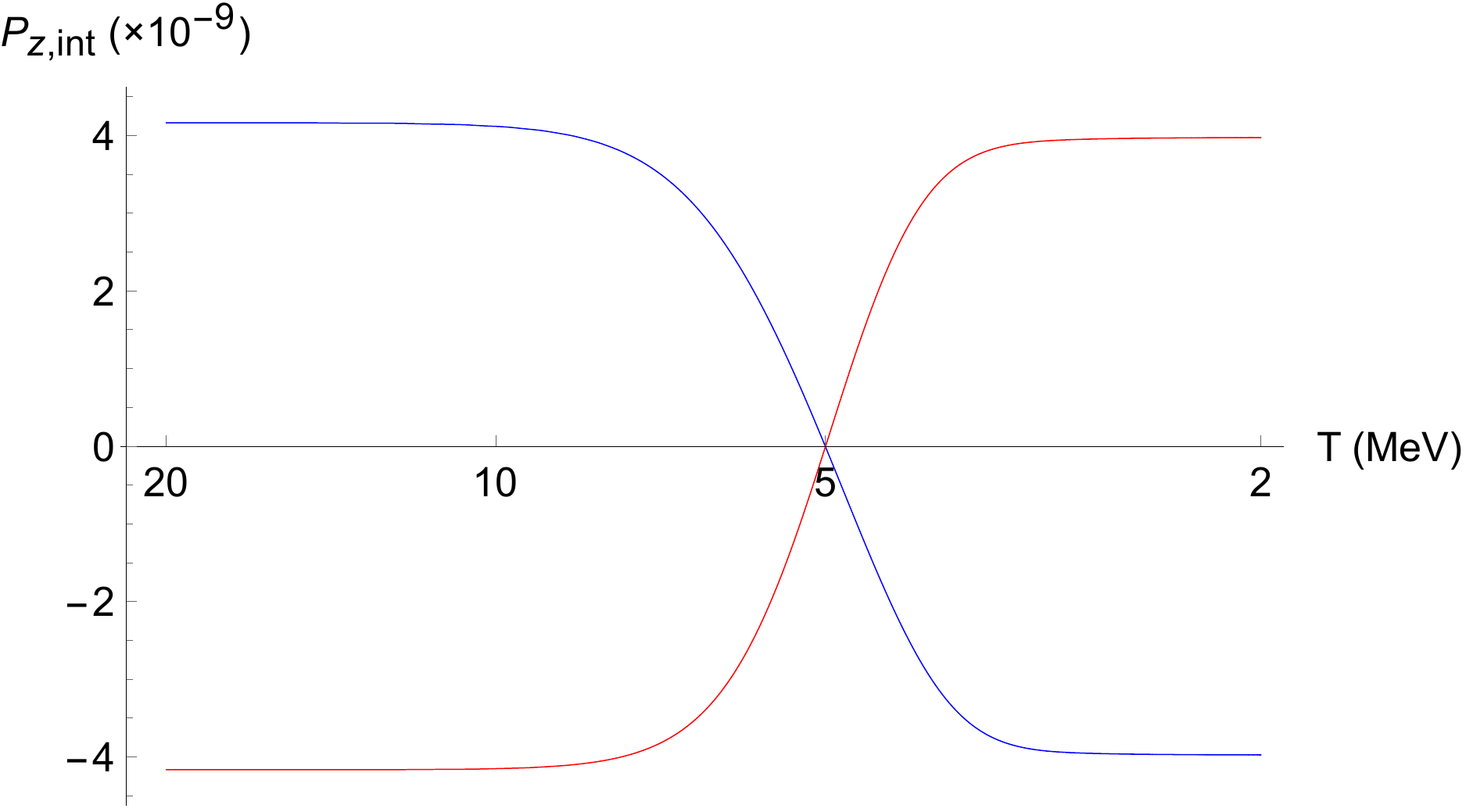}
\caption{(Color online) Symmetric MSW: $P_{z, \textrm{int}}$ (blue, upper curve at $T = 20$ MeV) and $\bar{P}_{z, \textrm{int}}$ (red) in the IH with initial degeneracy parameters $\eta_{\nu_e} = 5\times10^{-8}$, $\eta_{\nu_x} = 0$. \label{5e8intPz}}
\end{figure}

Generally speaking, the dominant feature in the flavor-transformation landscape is the equality of $| \mathcal{H}_{\textrm{vac}, z} |$ and $| \mathcal{H}_{e, z} |$, which for $1-3$ mixing occurs in the region of $T \sim 5$ MeV.  For $\eta_{\nu_e} \lesssim 5 \times 10^{-8}$ this is the \textit{only} feature (Fig.~\ref{5e8intPz}), as the self-coupling is so weak as to leave transformation through the resonance essentially untouched.  At these small lepton numbers --- as would be expected if neutrino--neutrino scattering were simply omitted --- neutrinos and antineutrinos of all modes undergo complete MSW conversion.  We emphasize that, unlike in a supernova environment, \textit{both} neutrinos and antineutrinos resonantly transform due to $\mathcal{H}_e$ being CP-symmetric: At high temperatures $\nu_e$ and $\bar{\nu}_e$ are at energies lower than $\nu_x$ and $\bar{\nu}_x$, respectively, thanks to the thermal potential, but at low temperatures (in vacuum) are at higher energies, thanks to the IH.

Evidently, if the neutrino chemical potential is entirely in $\nu_e$ and if it is of the same order as the baryon asymmetry $\eta$, then neutrino--neutrino scattering has an ignorable impact on flavor evolution.  This conclusion is unsurprising given Fig.~\ref{potentials}, which shows that $| \mathcal{H}_{\nu, z} |$ for $\eta_{\nu_e} = 5 \times 10^{-8}$ is always about an order of magnitude or more below either $| \mathcal{H}_{\textrm{vac}, z} |$ or $| \mathcal{H}_{e, z} |$. It is worth pointing out that at such small lepton numbers the $e^\pm$ finite-density potential $\mathcal{H}^\textrm{(FD)}_e = \sqrt{2} G_F \left( n_{e^-} - n_{e^+} \right) \mathrm{L}$ should be included in the equations of motion for consistency, but this term likewise makes an inconsequential contribution to the total Hamiltonian.  The effect of the thermal potential from the neutrino background is yet more feeble.

\subsubsection{$\eta_{\nu_e} = 1.5 \times 10^{-7}$: Partial MSW}

\begin{figure}
\includegraphics[width=.47\textwidth]{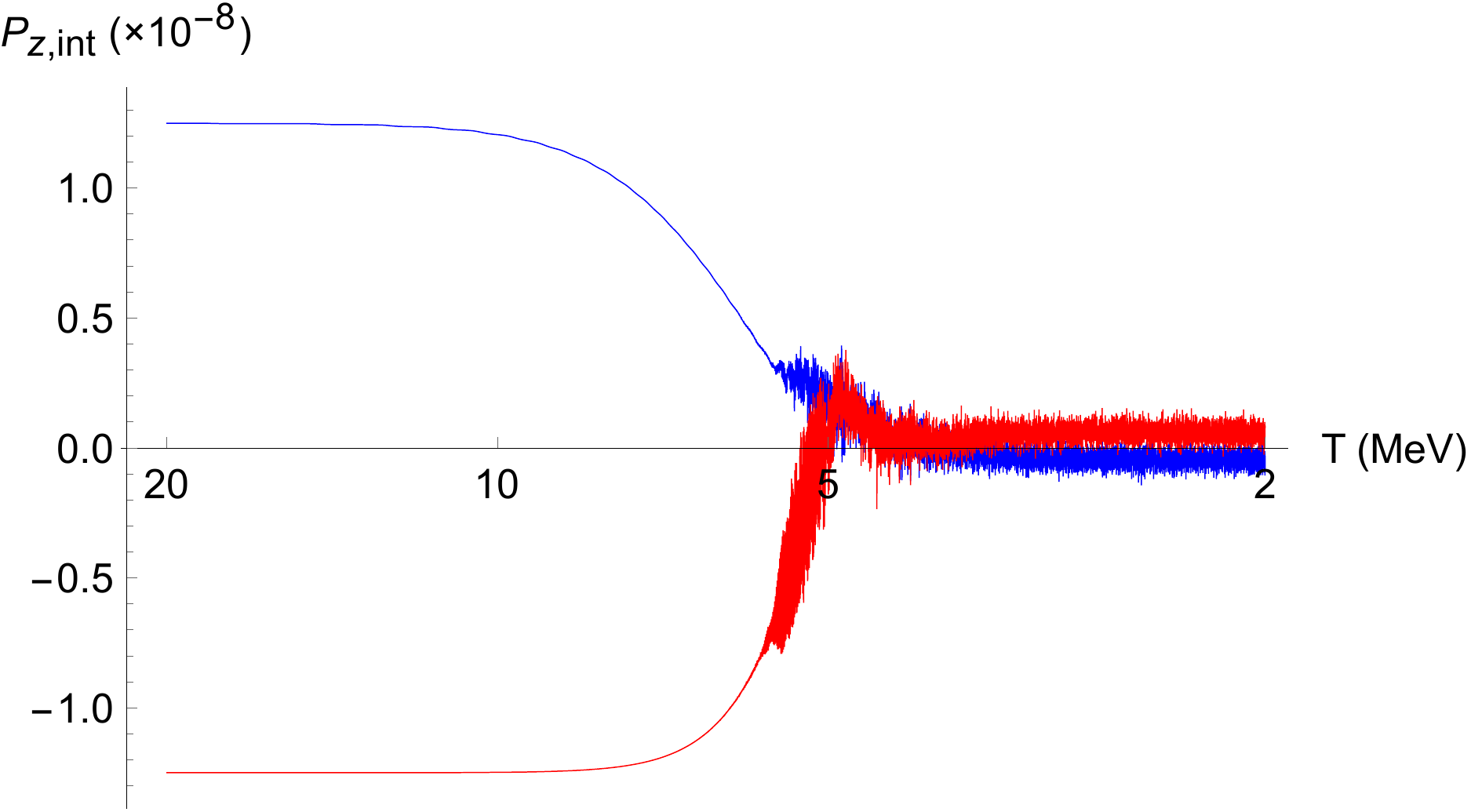}
\caption{(Color online) Partial MSW: $P_{z, \textrm{int}}$ (blue, upper curve at $T = 20$ MeV) and $\bar{P}_{z, \textrm{int}}$ (red) in the IH with initial degeneracy parameters $\eta_{\nu_e} = 1.5\times10^{-7}$, $\eta_{\nu_x} = 0$. \label{15e7intPz}}
\end{figure}

\begin{figure}
\includegraphics[width=.47\textwidth]{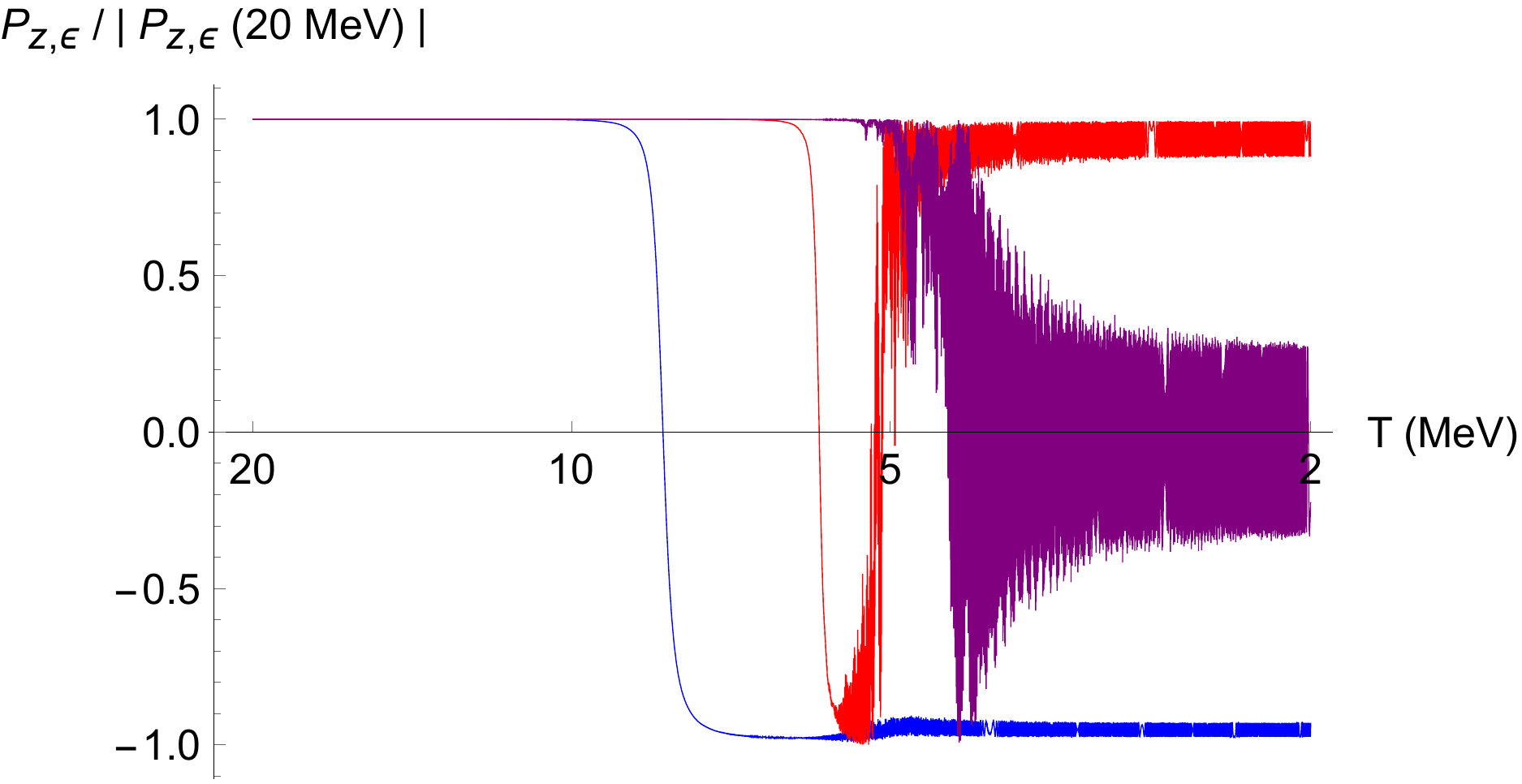}
\caption{(Color online) Partial MSW: $P_{z,\epsilon}$ for $\epsilon = 1.15$ (blue, bottommost curve at $T = 2$ MeV), $2.36$ (red, topmost curve at $T = 2$ MeV), and $4.78$ (purple), with the same parameters as in Fig.~\ref{15e7intPz}.  Note that here and in subsequent plots $P_{z,\epsilon}$ has been normalized to an initial value of unity for each $\epsilon$; this choice puts all modes on equal footing for the purpose of comparing flavor evolution.   \label{15e7indPz}}
\end{figure}

\begin{figure}
\includegraphics[width=.47\textwidth]{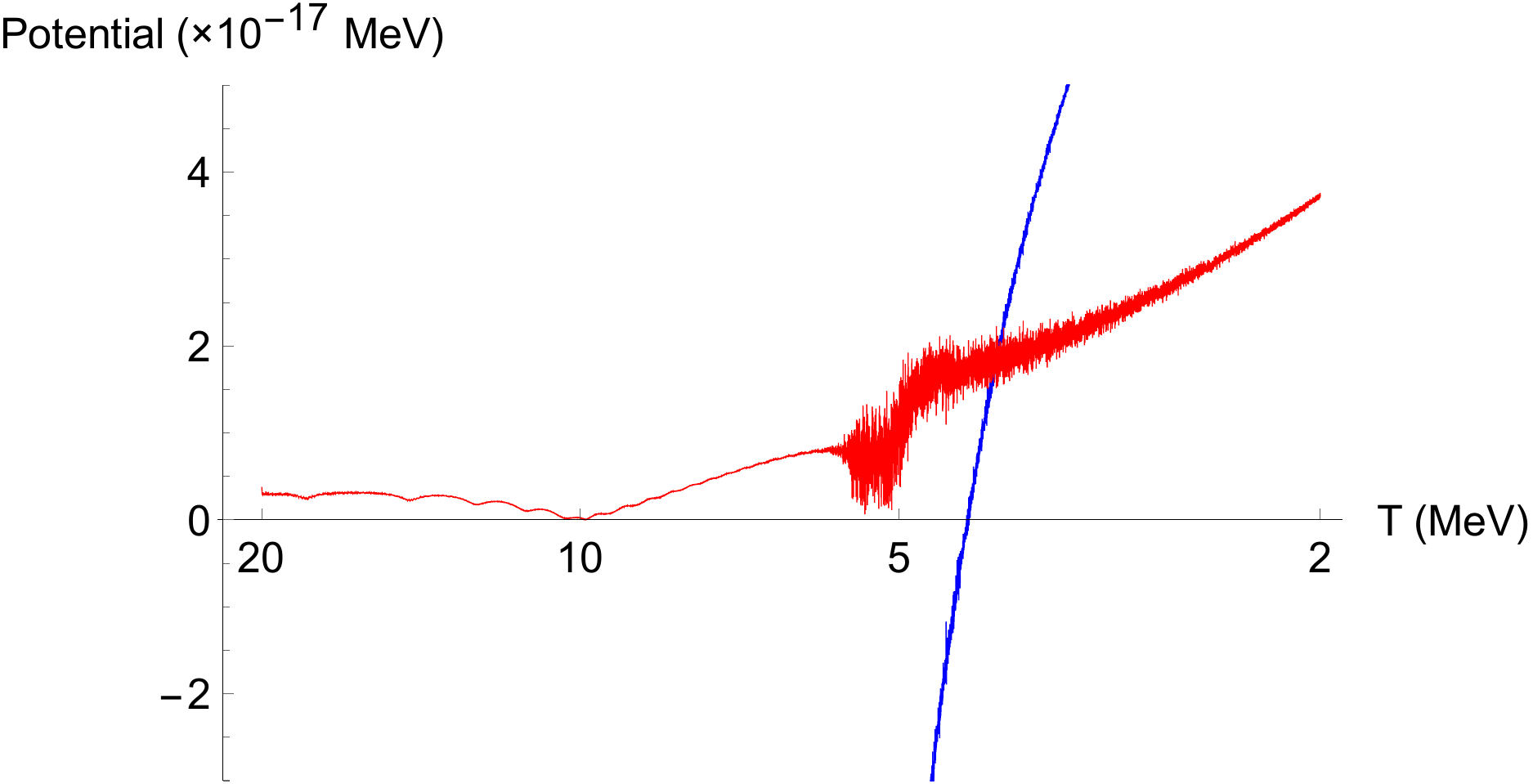}
\caption{(Color online) Partial MSW: $\mathcal{H}_z$ (blue, nearly vertical curve) and $| \mathcal{H}_T |$ (red) as functions of $T$ for the $\epsilon = 4.78$ mode shown in Fig.~\ref{15e7indPz}. \label{15e7neuN6H}}
\end{figure}

As $\eta_{\nu_e}$ is scaled up, MSW transformation becomes overall less effective for both neutrinos and antineutrinos (Fig.~\ref{15e7intPz}).  The incompleteness of the conversion of $P_{z, \textrm{int}}$ and $\bar{P}_{z, \textrm{int}}$ is attributable to the differing outcomes of individual modes: The lowest-energy modes go through MSW unfettered while higher-energy modes exhibit large, aperiodic oscillations of high frequency (Fig.~\ref{15e7indPz}).

The higher-energy modes transform inefficiently due to a loss of adiabaticity, as indicated in Fig.~\ref{15e7neuN6H}.  Prior to resonance the off-diagonal potential $\mathcal{H}_T$ fluctuates rapidly and achieves (nearly) vanishing magnitude at various points.  Since $| \mathcal{H}_T |$ mediates the transition probability between states, this behavior allows neutrinos to depart from their initial energy-eigenstate track at the level-crossing.  In Sec.~\ref{subadiabaticity} we introduce the quantitative measure of adiabaticity traditionally used in studies of resonant neutrino conversion, and we explore further the role it plays in our results.  For now, suffice it to say that in this regime self-coupling suppresses adiabaticity because it is strong enough to influence $\mathcal{H}$ but not strong enough to force the individual modes to pass through resonance collectively.  Such behavior is connected to the fact that for most modes the three contributions to the Hamiltonian are of comparable magnitude in the MSW region.

\subsubsection{$\eta_{\nu_e} = 2 \times 10^{-6}$: Asymmetric ($\nu$- or $\bar{\nu}$-only) MSW}

\begin{figure}
\includegraphics[width=.47\textwidth]{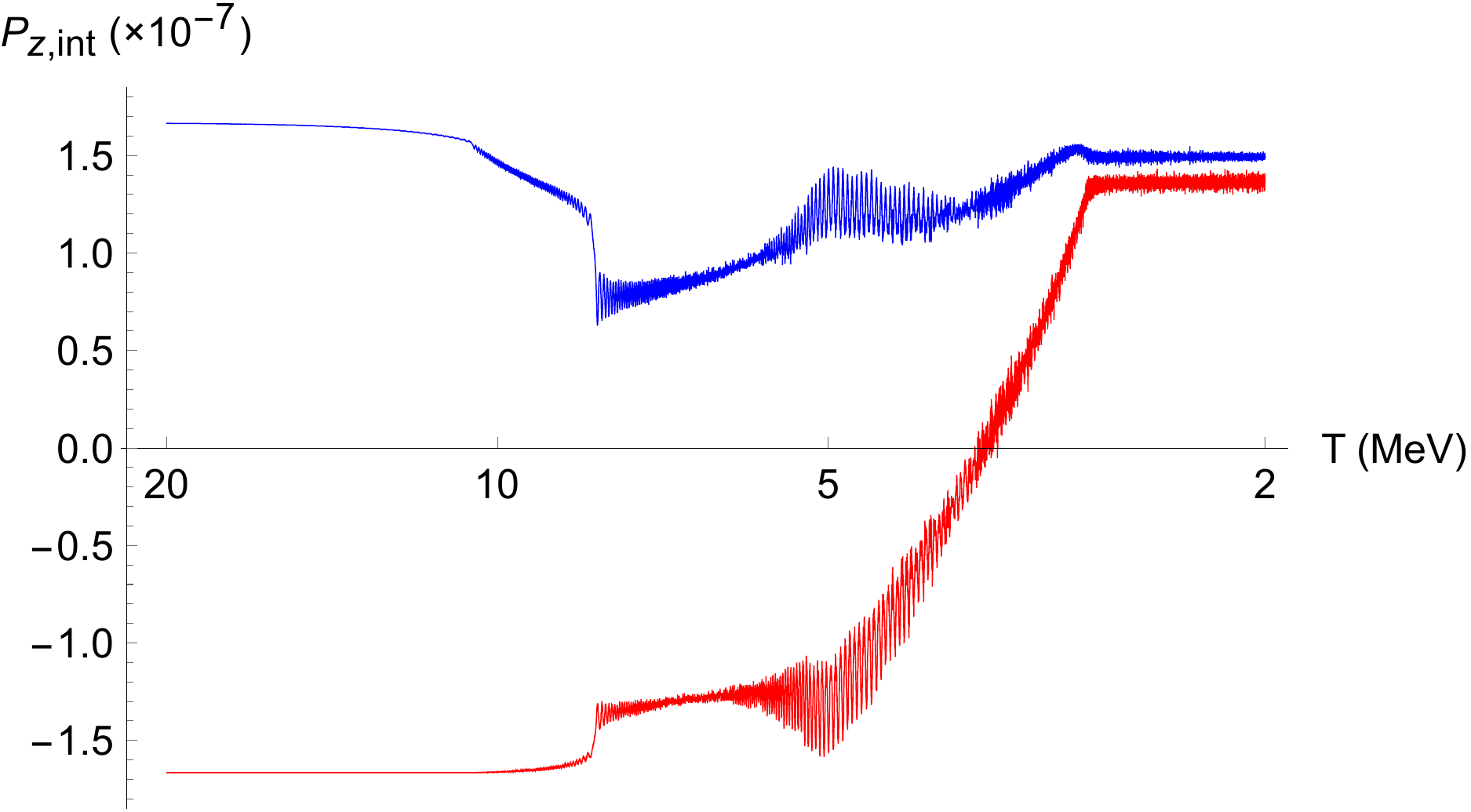}
\caption{(Color online) Asymmetric MSW: $P_{z, \textrm{int}}$ (blue, upper curve at $T = 20$ MeV) and $\bar{P}_{z, \textrm{int}}$ (red) in the IH with initial degeneracy parameters $\eta_{\nu_e} = 2\times10^{-6}$, $\eta_{\nu_x} = 0$. \label{2e6intPz}}
\end{figure}

Moving to greater values of $\eta_{\nu_e}$, the partial conversion of neutrinos becomes even more stunted while the conversion of antineutrinos actually grows more \textit{effective} (Fig.~\ref{2e6intPz}).

The cancellation of $\mathcal{H}_{\textrm{vac}, z}$ and $\mathcal{H}_{e, z}$ in the MSW region precipitates some degree of transformation in both neutrinos and antineutrinos.  However, since $\mathcal{H}_{\nu, z}$ exceeds the other two by a factor of $\sim10$ in magnitude, MSW conversion is stillborn (in the case of neutrinos) or delayed until the vacuum potential overtakes the self-coupling soon thereafter (in the case of antineutrinos).  Starting at the MSW region and continuing down to $T \sim 3$ MeV, antineutrinos gradually cross over from predominantly $\bar{\nu}_x$ to predominantly $\bar{\nu}_e$; by the bottom of this temperature range they have almost completely transformed.

Neutrinos undergo only marginal conversion because the large self-coupling potential, which enhances the effective mass of $\nu_e$ relative to $\nu_x$, props up $\nu_e$ into the higher energy eigenstate over most of this temperature range, thus wiping out what would otherwise be an MSW resonance.  (A level-crossing does occur at higher temperature where the thermal and self-coupling potentials cancel, but this resonance appears well before the MSW region and, as we will discuss in Sec.~\ref{mnrsec}, is neutralized by non-adiabaticity.)  Conversely, the initial population of $\bar{\nu}_x$ is effectively immersed in a bath of $\nu_e$, which serves to elevate the energy of $\bar{\nu}_x$ over that of $\bar{\nu}_e$ until $\mathcal{H}_\textrm{vac}$ becomes dominant.   Hence self-coupling does not eliminate the antineutrino level-crossing in the MSW region, though it does significantly alter evolution through it.

A notable characteristic of this regime is that the location of $| \mathcal{H}_{\nu, z} | \sim | \mathcal{H}_{\textrm{vac}, z} |$ has been pulled away from that of $| \mathcal{H}_{e, z} | \sim | \mathcal{H}_{\textrm{vac}, z} |$ --- compare to the partial MSW regime, where they coincide --- but the regions are still close enough together that the flavor transformation instigated by the traditional MSW mechanism can be capitalized on to enact a flavor swap by the later $| \mathcal{H}_{\nu, z} | \sim | \mathcal{H}_{\textrm{vac}, z} |$ cancellation.  As we observe in the next regime, increasing further the separation between the two locations leads to MSW manqu\'e --- but here the separation actually salvages efficient conversion of antineutrinos.

\subsubsection{$\eta_{\nu_e} = 5 \times 10^{-5}$: Minimal transformation}

\begin{figure}
\includegraphics[width=.47\textwidth]{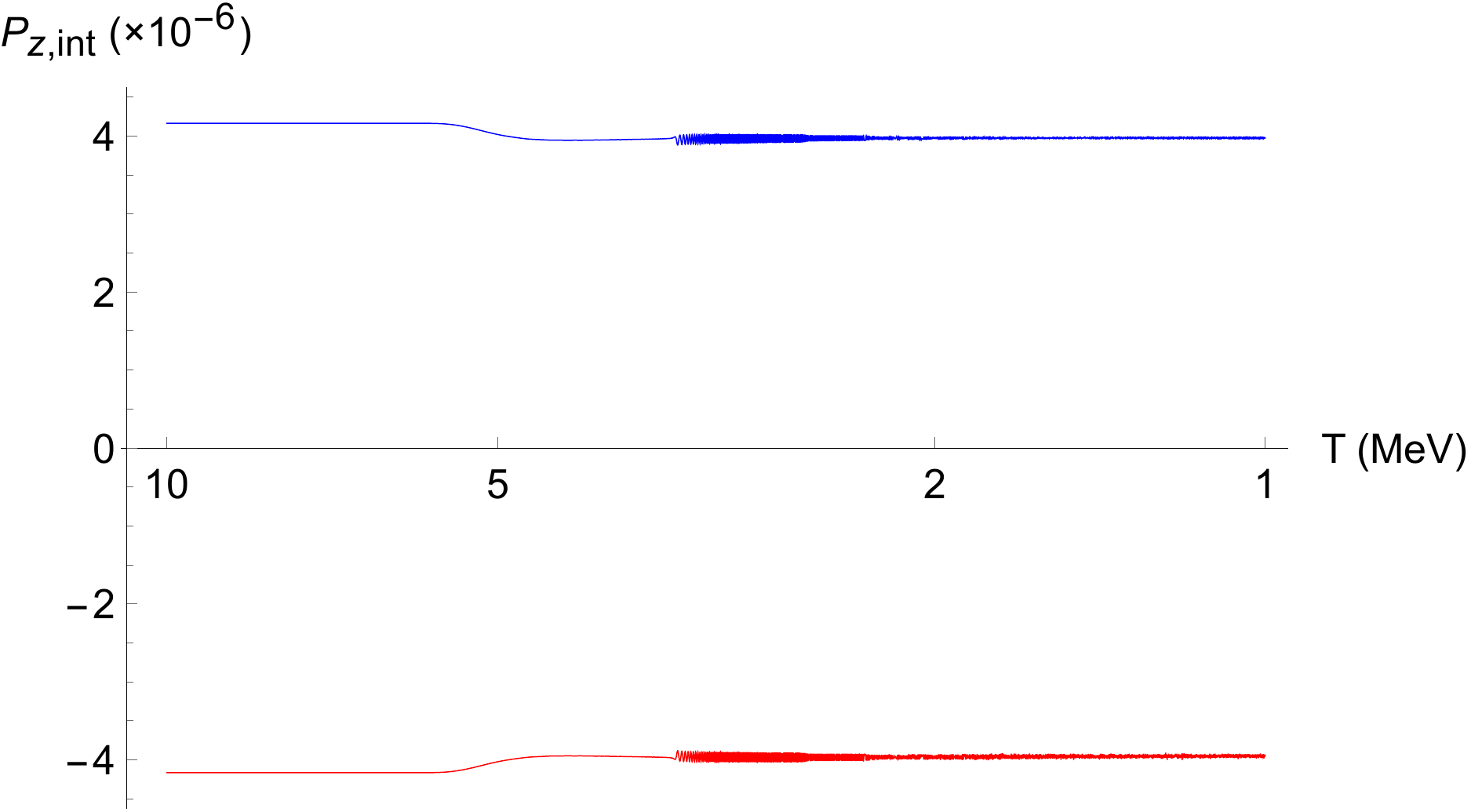}
\caption{(Color online) Minimal transformation: $P_{z, \textrm{int}}$ (blue, upper curve at $T = 10$ MeV) and $\bar{P}_{z, \textrm{int}}$ (red) in the IH with initial degeneracy parameters $\eta_{\nu_e} = 5\times10^{-5}$, $\eta_{\nu_x} = 0$. \label{5e5intPz}}
\end{figure}

With $\eta_{\nu_e} = 5 \times 10^{-5}$ the locations of $| \mathcal{H}_{\nu, z} | \sim | \mathcal{H}_{\textrm{vac}, z} |$ and $| \mathcal{H}_{e, z} | \sim | \mathcal{H}_{\textrm{vac}, z} |$ are well removed from one another.  As a result the MSW level-crossing is now thwarted entirely, and virtually no flavor conversion takes place (Fig.~\ref{5e5intPz}).  What transformation does occur commences near $T \sim 5$ MeV, as usual, but fails to get very far due to the strong ``inertial'' effect exerted by $\mathcal{H}_\nu$.  The self-coupling keeps $\nu_e$ and $\bar{\nu}_x$ in the heavier eigenstates throughout MSW, preventing $\mathcal{H}_z$ from ever crossing into negative territory (or $\bar{\mathcal{H}}_z$ into positive).

Since coherence between the flavors only marginally develops at these lepton numbers, there is meager fuel for decoherence to consume, and the minimal-transformation regime is consequently the best preserver of its initial lepton asymmetry when damping is turned on.  It is a tantalizing coincidence that this regime also encompasses the range of lepton numbers suggested by resonant production of sterile neutrino dark matter, which favors the neighborhood of $L_\nu \sim 5 \times 10^{-4}$ \cite{abazajian2014} when the $\sim 3.55$ keV X-ray line of Refs.~\cite{bulbul2014, boyarsky2014} is attributed to the decay of sterile neutrinos.  These lepton numbers occupy the top end of the minimal-transformation regime, where synchronized oscillations are beginning to grow in amplitude but are still unable to realize a large net conversion of flavor.

A phenomenon notably absent from this regime and the foregoing ones is the spectral swap, in which nearly all antineutrinos below a certain energy threshold change flavor and nearly all antineutrinos above the threshold do not (or similarly for neutrinos) \cite{duan2006, raffelt2007}.  For $\eta_{\nu_e} \lesssim 10^{-7}$ spectral swaps are ruled out by virtue of the fact that the self-coupling potential never dominates.  But for $\eta_{\nu_e} = 5 \times 10^{-5}$, for instance, $\mathcal{H}_\nu$ remains dominant sufficiently far below temperatures at which $| \mathcal{H}_{\textrm{vac},z} | \sim | \mathcal{H}_{e,z} |$ that the requisite conditions for a spectral swap might be thought to prevail as $\mathcal{H}_{\textrm{vac}, z}$ finally does overtake $\mathcal{H}_{\nu,z }$.  In actuality the spectral swap is preempted by the MSW region, which in the minimal-transformation regime deposits neutrinos and antineutrinos essentially into the nearest mass eigenstates.  With all well-populated modes already in mass eigenstates before $\mathcal{H}_\textrm{vac}$ takes over, no spectral swap can occur.  The synchronized-oscillation regime proves to be the exception to this trend, as we discuss below.

\subsubsection{$\eta_{\nu_e} = 5 \times 10^{-3}$: Large synchronized oscillations}

\begin{figure}
\includegraphics[width=.47\textwidth]{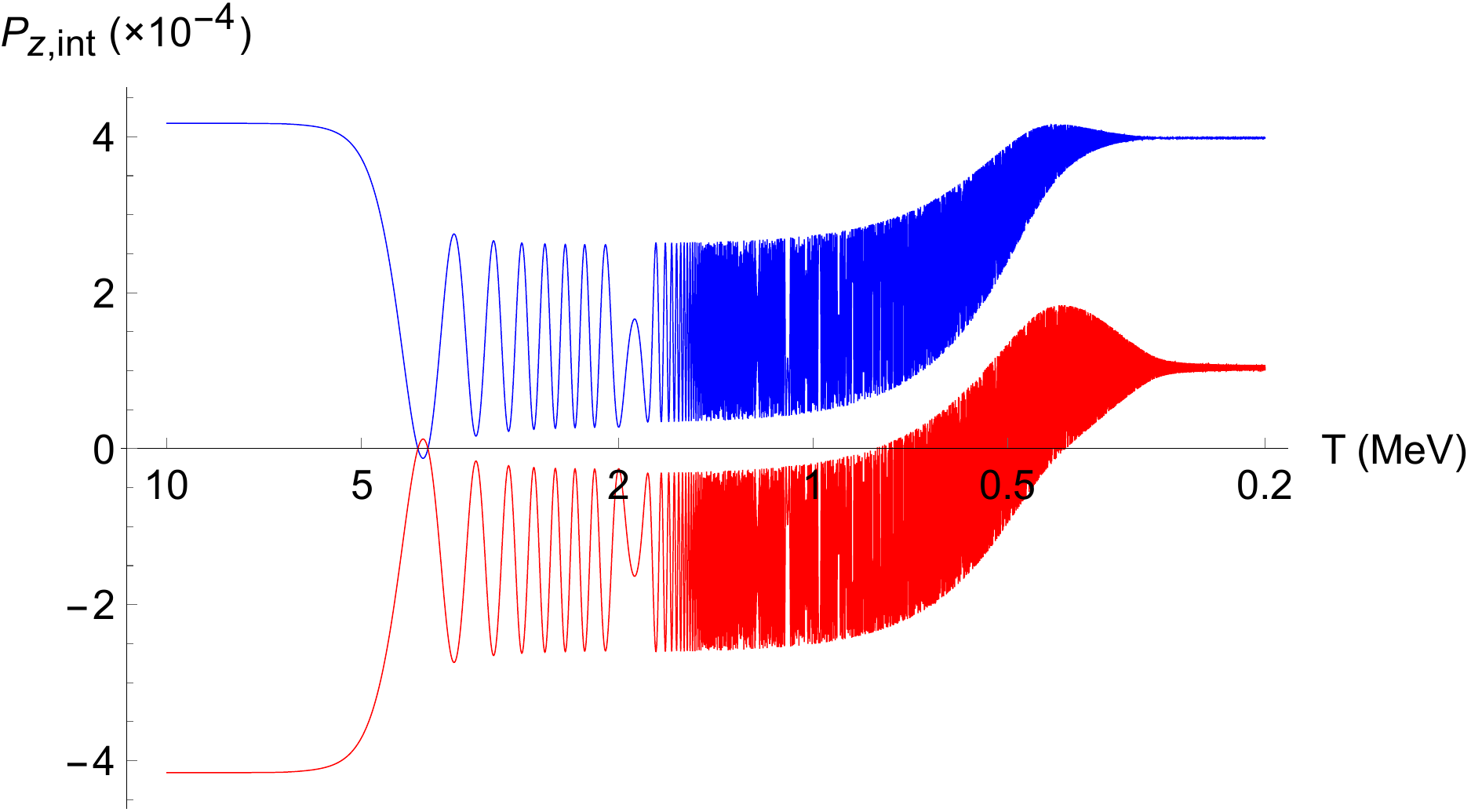}
\caption{(Color online) Large synchronized oscillations: $P_{z, \textrm{int}}$ (blue, upper curve at $T = 10$ MeV) and $\bar{P}_{z, \textrm{int}}$ (red) in the IH with initial degeneracy parameters $\eta_{\nu_e} = 5\times10^{-3}$, $\eta_{\nu_x} = 0$. \label{5e3intPz}}
\end{figure}

In this regime the lepton asymmetry is large enough that neutrino--neutrino scattering shifts towards promoting rather than resisting transformation.  Once the expansion rate and the $e^{\pm}$ density have dropped sufficiently, large synchronized oscillations ensue, with all of the modes locked together by self-coupling (Fig.~\ref{5e3intPz}).

Although on the face of it this regime hosts perhaps the most active flavor evolution, in some ways the behavior is just that of the minimal-transformation regime writ large.  In both regimes modes undergo synchronized oscillations after first gesturing towards MSW conversion, and then move into mass eigenstates as $\mathcal{H}_\nu$ becomes unimportant.  But for larger $\eta_{\nu_e}$ the gesture towards MSW is stronger, the synchronized oscillations last longer and have larger amplitudes, and the movement into mass eigenstates entails more significant transformation at late time. The minimal-transformation scenario of $\eta_{\nu_e} = 5 \times 10^{-5}$ is an extreme example of the shrinking of these features, down to a size indiscernible at the scale of Fig.~\ref{5e5intPz}.

\begin{figure}
\includegraphics[width=.47\textwidth]{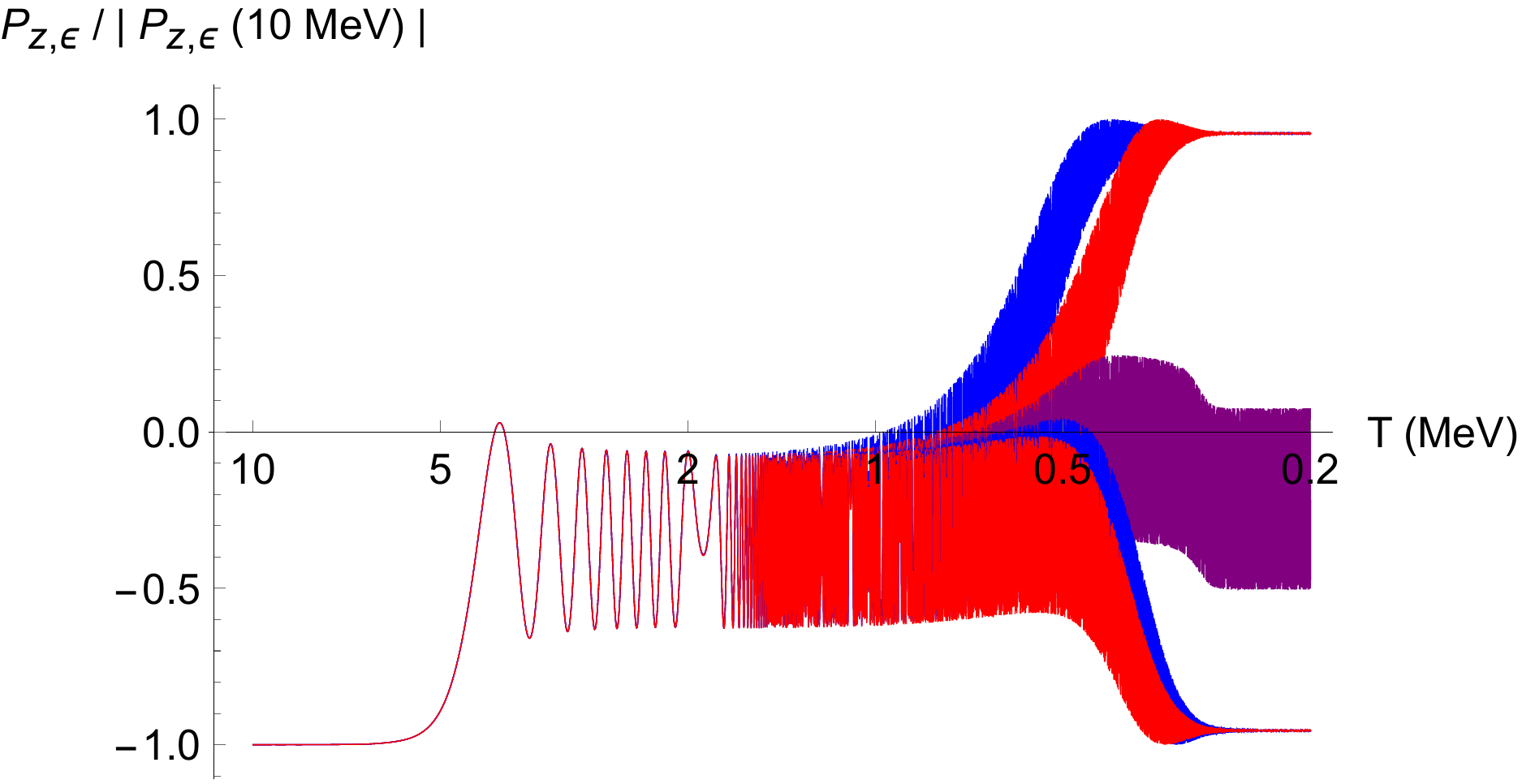}
\caption{(Color online) Large synchronized oscillations: $\bar{P}_{z,\epsilon}$ for $\epsilon = 1.15$ (upper blue curve at $T = 0.2$ MeV), 2.36 (upper red curve at $T = 0.2$ MeV), 3.57 (purple), 4.78 (lower blue curve at $T = 0.2$ MeV), and 5.99 (lower red curve at $T = 0.2$ MeV), computed with the same parameters as in Fig.~\ref{5e3intPz}. A spectral swap --- wherein modes below the threshold $\epsilon_\textrm{th} \approx 3.5$ change flavor and those above do not --- is evident.  \label{5e3swap}}
\end{figure}

The qualitatively novel feature that distinguishes the synchronized-oscillation regime from the minimal-transformation regime is that (for $\eta_{\nu_e} > 0$) antineutrinos do not return en masse to the lighter mass eigenstate; instead many modes move to the heavier one, more closely associated with $\bar{\nu}_e$.  Conversion of antineutrinos in this manner is more dramatic for larger initial $\eta_{\nu_e}$, even causing $\bar{P}_{z, \textrm{int}}$ to change sign for $\eta_{\nu_e} \gtrsim 5 \times 10^{-3}$.  The upward drifting of $P_{z, \textrm{int}}$ at low temperatures reflects the spectral swap that occurs as $\mathcal{H}_\textrm{vac}$ comes to dominate (Fig.~\ref{5e3swap}).  The threshold energy, below which $\bar{\nu}$ swap, moves up to higher $\epsilon$ as the lepton asymmetry is increased; it is for this reason that the spectral swap has no discernible impact on the minimal-transformation regime.

Large-amplitude synchronized oscillations are associated with a solution of the equations of motion in which the off-diagonal elements of $\mathcal{H}_\nu$ steer the evolution of the system into self-sustained maximal mixing for both neutrinos and antineutrinos \cite{fuller2006}.  What our results highlight is the fact that this solution is not easily accessed in the early universe: As shown in Fig.~\ref{5e3intPz}, even an asymmetry of $\sim 10^{-3}$ does not foster maximal mixing, even though the mixing angle is still significantly enhanced over its value in vacuum.  In the minimal-transformation regime, where the mixing angle is \textit{suppressed}, the failure to enter this off-diagonal-driven mode is at its most spectacular.

\begin{figure}
\includegraphics[width=.47\textwidth]{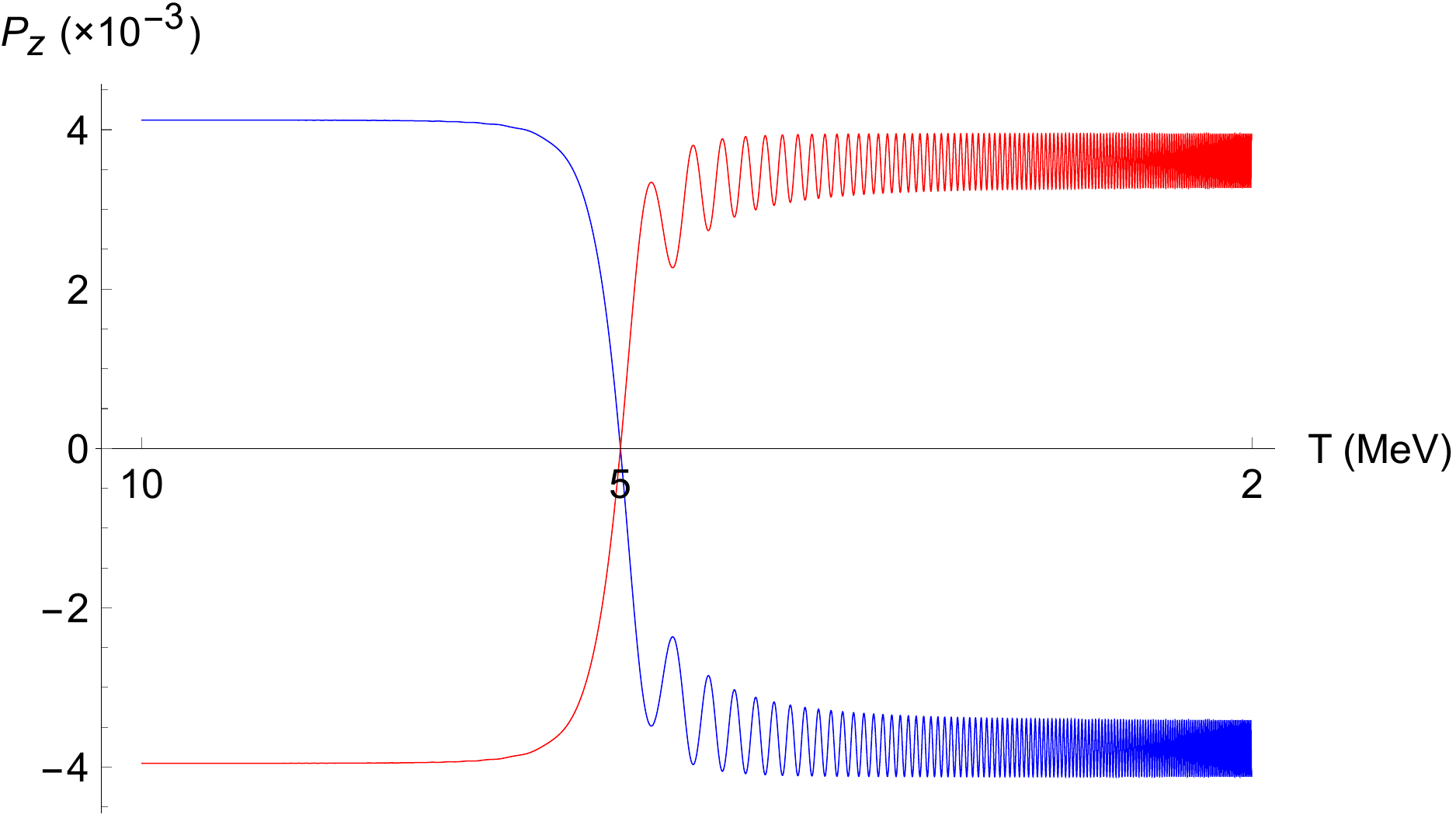}
\caption{(Color online) Top end of the synchronized-oscillation regime: $P_{z, \textrm{int}}$ (blue, upper curve at $T = 10$ MeV) and $\bar{P}_{z, \textrm{int}}$ (red) in the IH with initial degeneracy parameters $\eta_{\nu_e} = 5\times10^{-2}$, $\eta_{\nu_x} = 0$. \label{5e2intPz}}
\end{figure}

In a sense the very largest allowable lepton asymmetries --- those about an order of magnitude greater even than the exemplar asymmetry portrayed in Fig.~\ref{5e3intPz} --- actually overshoot this mode of self-sustained maximal mixing, displaying instead synchronized MSW transformation at $T \sim 5$ MeV followed by synchronized oscillations of non-maximal amplitude (Fig.~\ref{5e2intPz}).  The phenomenon of synchronized MSW, where all modes undergo efficient MSW conversion in unison, can be understood from the following perspective.  Decomposing the Hamiltonian into its constituents and taking the coherent limit, Eq.~\eqref{veceom} becomes
\begin{equation}
Hx \frac{d\vec{P}}{dx} = \left( \vec{\mathcal{H}}_\textrm{vac} + \vec{\mathcal{H}}_e + \vec{\mathcal{H}}_\nu \right) \times \vec{P}. \label{veceomconst}
\end{equation}
At high temperature all of the individual modes point along the $z$-axis, and in the limit $| \vec{\mathcal{H}}_\nu | \gg | \vec{\mathcal{H}}_\textrm{vac} |$, $| \vec{\mathcal{H}}_e |$ they remain locked together even as the temperature cools and they depart from that axis.  Their alignment implies that, for any given mode, $\vec{P}$ (very nearly) points along $\vec{\mathcal{H}}_\nu$ and so Eq.~\eqref{veceomconst} can be approximated as
\begin{equation}
Hx \frac{d\vec{P}}{dx} \approx \left( \vec{\mathcal{H}}_\textrm{vac} + \vec{\mathcal{H}}_e \right) \times \vec{P}. 
\end{equation}
The upshot is that all modes follow the track that the average energy mode $\epsilon \approx 3.15$ would undergo if there were \textit{no} self-coupling.  (Further details on synchronized MSW conversion are provided in, for example, Ref.~\cite{abazajian2002}.)

Lepton asymmetries at the top end of the synchronized-oscillation regime are converging on this limit, but as shown in Fig.~\ref{5e2intPz} --- for an initial degeneracy parameter $\eta_{\nu_e} = 5 \times 10^{-2}$ --- the resonant conversion is incomplete, as the approximation that all $\vec{P} \left( \epsilon \right)$ are aligned is an imperfect one.  Since $| \vec{\mathcal{H}}_{\nu} | \gg | \vec{\mathcal{H}}_\textrm{vac} |$ for the entire temperature range depicted in Fig.~\ref{5e2intPz}, synchronized oscillations then take over at $T \lesssim 5$ MeV, once $| \vec{\mathcal{H}}_e |$ has fallen off.  As the lepton asymmetry is dialed up further, the efficiency of conversion through the synchronized MSW mechanism increases and the amplitude of post-MSW synchronized oscillations decreases.  In a somewhat poetic turn, the evolution of $P_{z, \textrm{int}}$ and $\bar{P}_{z, \textrm{int}}$ at infinite lepton asymmetry is identical (up to scale) to that at zero lepton asymmetry.

We wish to underscore the point that despite the dominance by several orders of magnitude of $\mathcal{H}_\nu$ all the way through the MSW region, this regime strongly bears the fingerprints of the matter background.  If it were not for the cancellation between $\mathcal{H}_{\textrm{vac}, z}$ and $\mathcal{H}_{e, z}$, the amplitude of the oscillations would be diminished down to the scale set by the vacuum mixing angle (as indeed it is in the NH), and the spectral swaps at these lepton asymmetries would be erased.  The synchronized-oscillation regime thus highlights the insistent influence that can be exerted even by a \textit{would-be} MSW resonance.

\subsection{Adiabaticity \label{subadiabaticity}}

In our discussion of the five regimes just laid out, we have stressed the decisive role of level-crossings in determining flavor transformation.  But the presence or absence of level-crossings is not the whole story.  An important tool for understanding the behavior of neutrinos as they pass through resonance is the adiabaticity parameter $\gamma$, which quantifies the efficiency of flavor conversion \cite{bethe1986, haxton1986, parke1986, kim1987}.  The parameter is defined as
\begin{equation}
\gamma \equiv 2\pi \frac{\delta t}{l_M^\textrm{res}} \approx \Delta_M^\textrm{res} \left| \frac{d\mathcal{H}_z}{dt} \right|^{-1}_{\textrm{res}} \delta \mathcal{H}_z, \label{gammadef}
\end{equation}
with $l_M^\textrm{res} \equiv 2 \pi / \Delta_M^\textrm{res}$ the in-medium oscillation length at resonance and $\delta t$ the resonance width, which is to say the time required for $\sin^2 2\theta_M$ to fall to half its resonant value.  The approximation above comes from the definition of $l_M^\textrm{res}$ and a recasting of $\delta t$ in terms of $\delta \mathcal{H}_z$.  Since the self-coupling and thermal potentials are varying much more rapidly than the vacuum potential, we can make the further approximation that, for the purposes of computing adiabaticity, $\mathcal{H}_{\textrm{vac}, z}$ is constant.  We then obtain an expression for $\gamma$ equivalent to that in Ref.~\cite{abazajian2001b}.

An adiabaticity parameter $\gamma \gg 1$ corresponds to a resonance width broad enough to contain many oscillation lengths, indicating that the potentials change sufficiently slowly that neutrinos are able to track the Hamiltonian through the level-crossing.  A small value of $\gamma$, conversely, corresponds to a large probability of neutrinos jumping from one energy eigenstate to the other: The Landau--Zener probability for such a transition is $P \approx e^{-\pi \gamma /2}$ \cite{landau1932, zener1932}.  The early universe is ripe for adiabaticity, as $\gamma$ is ultimately a comparison of the fast-fluttering dynamical timescale set by oscillations to the molasses-like Hubble timescale set by gravity.  We will see, however, that under certain circumstances self-coupling can compromise this propensity.

Resonance occurs whenever the vacuum potential cancels with the weak-interaction potential, producing degenerate instantaneous energy eigenstates:
\begin{equation}
\mathcal{V}_z = \frac{\delta m^2 \cos 2\theta}{2 \epsilon_\textrm{res} T}.
\end{equation}
The left-hand side implicitly depends on $\epsilon_\textrm{res}$.  (Recall the definition of $\mathcal{V}$ in and below Eq.~\eqref{hvmats}.)  Solving for the resonant comoving energy yields
\begin{equation}
\epsilon_\textrm{res} = \frac{\mathcal{H}_{\nu, z}}{2 \tilde{\mathcal{H}}_{e, z}} \left( 1 \pm \sqrt{1 - \frac{2 \delta m^2 \cos 2\theta}{T} \frac{\tilde{\mathcal{H}}_{e, z}}{(\mathcal{H}_{\nu, z})^2} } \right), \label{epsresgeneral}
\end{equation}
with $\tilde{\mathcal{H}}_{e, z} = | \mathcal{H}_{e, z} | / \epsilon$.  While this expression is always valid, its predictive power, in the sense of allowing one to identify where resonance will occur without solving the equations of motion, is questionable due to the nonlinearity inherent in neutrino evolution.  Broadly, Eq.~\eqref{epsresgeneral} can be used to predict the locations of level-crossings only so long as no significant flavor transformation has yet occurred.  But once the polarization vectors have departed appreciably from their initial alignment along the $z$-axis, $\mathcal{H}_\nu$ has therefore also departed appreciably from its initial value, and so all bets are off as far as Eq.~\eqref{epsresgeneral} goes.  These comments are especially germane to the entire minimal-transformation regime and to much of the synchronized-oscillation regime, wherein resonance is never achieved despite the appearance that Eq.~\eqref{epsresgeneral} would countenance the existence of one.

Our numerical results demonstrate that tuning the lepton asymmetry does not considerably shift the location of the MSW resonance, provided that the self-coupling is not large enough to eliminate the resonance altogether.   This finding suggests that an analysis of the adiabaticity neglecting $\mathcal{H}_\nu$ may prove enlightening as to how the lepton asymmetry ``perturbs'' the matter-only MSW scenario.  Ignoring the contribution from self-coupling, the resonant weak-interaction potential is
\begin{equation} \mathcal{V}_z = \mathcal{H}^\textrm{res}_{e, z} = \left( \frac{7 \sqrt{2} \pi^2 G_F }{45 m^2_W} \left| \delta m^2 \right| \cos 2\theta \right)^{1/2} T^2, \end{equation}
introducing the notation $\mathcal{H}^\textrm{res}_{e, z}$ to denote the thermal potential of the mode instantaneously at resonance.  While for any particular mode $\epsilon$ the thermal potential $\mathcal{H}_{e, z}$ is dropping precipitously as $T^5$, the resonant thermal potential $\mathcal{H}^\textrm{res}_{e, z}$ drops only as $T^2$.  It turns out that the relatively sluggish descent of $\mathcal{H}^\textrm{res}_{e, z}$ ensures that MSW is always adiabatic in the early universe, so long as electrons and positrons are relativistic and the neutrino self-coupling can be neglected.  Under these circumstances the adiabaticity parameter is
\begin{equation}
\gamma \approx  \frac{1}{2^{3/4}} \frac{1}{5} \sqrt{\frac{7}{\pi}} \frac{m_{Pl}}{m_W} \sqrt{\frac{G_F}{g_*} | \delta m^2 | \cos 2 \theta \tan ^4 2 \theta }, \label{gammavt}
\end{equation}
where $m_{Pl}$ is the Planck mass, $m_W$ is the W boson mass, and $g_*$ is the number of relativistic degrees of freedom.  Both $m_{Pl}$ and $g_*$ enter through the derivative of the thermal potential, which is dictated by Hubble expansion: In the radiation-dominated epoch the Hubble constant is
\begin{equation}
H = \sqrt{\frac{8 \pi^3 g_*}{90}} \frac{T^2}{m_{Pl}}.
\end{equation}
We have taken $g_*$ to be constant over the span of temperatures relevant to this study, thus ignoring the small decrease that occurs as the last remaining $\mu^\pm$ disappear near the top of this temperature range and the later decrease that occurs as the $e^\pm$ population starts to become non-relativistic near the bottom.

Looking at Eq.~\eqref{gammavt}, the adiabaticity parameter is evidently independent of temperature when $\mathcal{H}_\nu = 0$ and is, moreover, very large: $\gamma \approx 130$ for $1-3$ mixing, guaranteeing that all modes undergo efficient MSW conversion, regardless of the temperature at which their respective resonances occur.  This fortuitous behavior, which is peculiar to the thermal potential, occurs because the resonance width and the in-medium oscillation length are growing with the same dependence on $T$.  The growth of the resonance width can be traced directly to the slowing-down of the Hubble rate $H \propto T^2$.

\begin{figure}
\includegraphics[width=.47\textwidth]{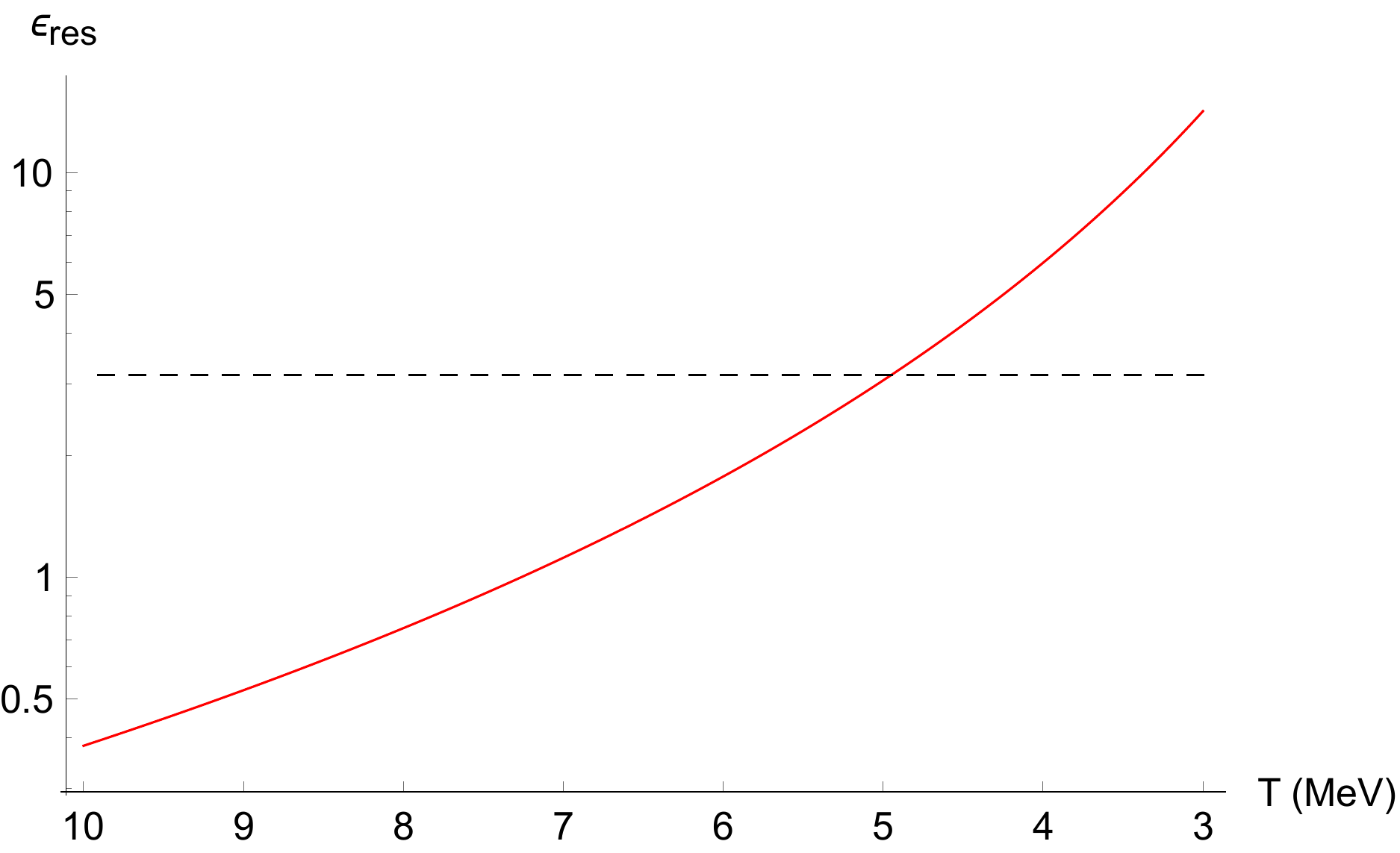}
\caption{(Color online) Resonant comoving energy $\epsilon_\textrm{res}$ (red, solid) as a function of T, with $\mathcal{H}_\nu = 0$.  Also plotted is the Fermi--Dirac average-energy mode $\epsilon_\textrm{avg} \approx 3.15$ (black, dashed). \label{eres}}
\end{figure}

Even as the resonance width is broadening, the rate at which the resonance sweeps upward through the energy modes is accelerating as a function of temperature: $\epsilon_\textrm{res} \propto 1 / T^3$.  Fig.~\ref{eres} shows $\epsilon_\textrm{res}(T)$ for $\mathcal{H}_\nu$ set to zero.  Reassuringly, $\epsilon_\textrm{avg} \approx 3.15$ becomes resonant right near $5$ MeV.

The preceding discussion gives credence to the notion that the resonant flavor transformation seen at $T \sim 5$ MeV across a range of lepton asymmetries is adiabatic by default --- that is, when only $\mathcal{H}_\textrm{vac}$ and $\mathcal{H}_\textrm{e}$ are considered.  But as our numerical results have revealed, self-coupling can obstruct the efficiency of resonant conversion in non-trivial ways.

One can glean some general insights into the effects of neutrino--neutrino scattering by re-deriving the in-medium mixing angle and mass-squared splitting, allowing in particular for the off-diagonal elements of $\mathcal{H}_\nu$.  In general these elements consist of nonzero real and imaginary parts, which (in keeping with our notation in Eq.~\eqref{hvmats}) we write as $\mathcal{V}_x$ and $\mathcal{V}_y$, respectively.  A complex potential, however, spoils the reformulation in terms of effective in-medium oscillation parameters, so we rotate to a flavor-space coordinate system in which the off-diagonal part of the entire Hamiltonian $\mathcal{H}$ is real.  Effective mixing parameters can be defined in this new coordinate system and then translated back in terms of $\mathcal{V}_x$ and $\mathcal{V}_y$ from the original, with the results
\begin{align} 
\Delta^2_M &= \left( \mathcal{V}_z - \Delta \cos 2\theta \right) ^2 + \left( \Delta \sin 2\theta + \mathcal{V}_x \right) ^2 + \mathcal{V}^2_y  \notag \\ 
\sin ^2 2\theta_M &= \frac{\left( \Delta \sin 2 \theta + \mathcal{V}_x \right) ^2 + \mathcal{V}^2_y}{\left( \Delta \sin 2 \theta + \mathcal{V}_x \right) ^2 + \mathcal{V}^2_y + \left( \mathcal{V}_z - \Delta \cos 2 \theta \right) ^2}. \label{instantmix}
\end{align}
It is important to note that these are only $\textit{instantaneous}$ mixing parameters, as the coordinate system required to make the off-diagonal elements of $\mathcal{H}$ real is constantly changing.  The validity of employing such a technique in an analysis of adiabaticity is made plausible by noting that rotations about the flavor axis do not mix $\mathcal{H}_z$ and $\mathcal{H}_T$.

Working from Eq.~\eqref{instantmix}, the resonance width expressed as a weak-interaction potential is
\begin{equation}
\delta \mathcal{V}_z = \sqrt{\left( \Delta \sin 2 \theta + \mathcal{V}_x \right) ^2 + \mathcal{V}^2_y},
\end{equation}
and, just as in the $\mathcal{H}_\nu = 0$ case, $\Delta_M^\textrm{res} = \delta \mathcal{V}_z$.  The definition of $\gamma$ (Eq.~\eqref{gammadef}) then leads to
\begin{align}
\gamma & \approx \frac{\left( \Delta \sin 2 \theta + \mathcal{V}_x \right)^2 + \mathcal{V}_y^2}{\left| 5 H \mathcal{H}_{e, z} + 3 H \mathcal{H}_{\nu, z} - \frac{\dot{L}_{\nu_e} - \dot{L}_{\nu_x}}{L_{\nu_e} - L_{\nu_x}} \mathcal{H}_{\nu, z} \right|_\textrm{res}}  \notag \\
& \approx ~~~~~~~~~~~~~~~~~ \left| \frac{\left| \mathcal{H}_T \right|^2}{\dot{\mathcal{H}}_z} \right|_\textrm{res}, \label{gammafull}
\end{align}
where in the last expression we emphasize an alternative interpretation of the adiabaticity parameter as the ratio of the off-diagonal part of $\mathcal{H}$ (squared) to the rate of change of its diagonal part.  In evaluating the derivative we have again taken $g_*$ to be constant.

While the term proportional to $\mathcal{H}_{e, z}$ in the denominator of Eq.~\eqref{gammafull} is always negative, the two terms proportional to $\mathcal{H}_{\nu, z}$ are of the same sign leading into resonance.  When $\mathcal{H}_{\nu, z}$ dominates over $\mathcal{H}_{e, z}$, the final term in the denominator therefore makes $\gamma$ smaller.  On the other hand, when $\mathcal{H}_{e, z}$ dominates, the term can either make $\gamma$ smaller (if $\eta_{\nu_e} < 0$ initially) or make it larger (if $\eta_{\nu_e} > 0$ initially).  That adiabaticity plummets as flavor conversion proceeds in the moderate-to-large-$L$ regime is well known from studies of resonant production of sterile neutrinos.  It reflects the fact that as the potential sweeps through resonance more rapidly, the resonance width contracts and conversion becomes less efficient.  What is less familiar is that flavor conversion can evidently feed back \textit{positively} on the adiabaticity of the resonance when the lepton number is small but nonzero.

Eq.~\eqref{gammafull} is comparable to expressions in Refs.~\cite{abazajian2001b, abazajian2005, kishimoto2006, kishimoto2008}, all of which consider resonant transformation between an active and a sterile state.  In that context the derivative of the lepton number drags down the adiabaticity with such resolve that the depletion of the lepton number ultimately halts the conversion process.  As suggested in the preceding paragraph, in our context as well the possibly adverse effect of $\dot{L}_{\nu_e}$ on $\gamma$ implies that adiabaticity may fail for some initial lepton asymmetries.  It deserves emphasis, however, that there is a crucial difference between the resonant production of sterile neutrino dark matter and the resonant conversion between active flavors: Because the sterile flavor eigenstate is uncharged under weak interactions, the forward-scattering neutrino--neutrino potential in an active--sterile system does not have off-diagonal elements.  In the polarization-vector picture for active--sterile mixing, the self-coupling potential consequently points along the $z$-axis, whereas for active--active mixing it tracks the polarization vectors away from the flavor axis.  This distinction corresponds in Eq.~\eqref{gammafull} to $\mathcal{V}_x$ and $\mathcal{V}_y$ being nonzero; it adds, as a result, another lever controlling adiabaticity.  In cases where cancellation occurs in the term $\left( \Delta \sin 2\theta + \mathcal{V}_x \right)^2$, the off-diagonal weak-interaction potential can in fact enfeeble $\gamma$, producing non-adiabaticity so long as $\mathcal{V}_y$ is not too large.  In other cases, though, off-diagonal self-coupling bolsters $\gamma$ by enlarging the resonance width and the in-medium mass-squared splitting.

Adiabaticity accounts for the general behavior seen in our numerical results wherever a level-crossing is present.  Despite these successes, as an analytical tool it has two shortcomings: One, it is too coarse an instrument to explain the precise evolution of \textit{individual} modes through resonance, which often display radically different behavior from one another even when nearby in energy.  (The partial-MSW regime exemplifies this point, as flavor evolution in this case exhibits highly non-trivial dependence on neutrino energy.)  And two, adiabaticity offers no insights into those regimes where the nonlinearity of self-coupling causes the system to avert resonance altogether.

\subsection{Matter--neutrino resonances in the early universe \label{mnrsec}}

In this paper we have presented scenarios in which flavor evolution prior to $T \sim 10$ MeV is quite restrained: The large potentials at high temperatures ensure that $\theta_M$ is minuscule, thereby preventing significant transformation away from the initial flavor eigenstates.  In truth it is not obvious \textit{a priori} that this statement always holds, as lepton asymmetries for which self-coupling dominates at $\sim 10$ MeV will have some higher temperature at which $\mathcal{H}_{e, z}$ surpasses $\mathcal{H}_{\nu, z}$ in magnitude.  If the two potentials are of opposite sign, then there will be a level-crossing at this higher, pre-MSW temperature.  Such a level-crossing has been dubbed a matter--neutrino resonance (MNR) and in recent years has been shown to be a possible conduit for significant flavor transformation in merger and accretion-disk environments \cite{malkus2012, malkus2014, malkus2016, vaananen2016, wu2016, zhu2016}.  (Related, albeit distinct, analyses have also been performed for supernovae \cite{qian1995a, *qian1995b}.) No explorations of MNR in the early universe have yet been conducted.

We have searched the regimes discussed above for signs of flavor transformation associated with the MNR mechanism.  Our numerical results have confirmed that level-crossings do indeed occur, but in the scenarios we have examined the transformation associated with these resonances is in most cases negligible.  The explanation appears to lie in the fact that the resonances are generally traversed non-adiabatically.  Referring again to Eq.~\eqref{gammafull}, $\gamma$ may be diminished either by a small off-diagonal potential $\mathcal{H}_T$ or by a large sweep rate in the diagonal potential $\mathcal{H}_z$.  We speculate that both factors are at play in preventing efficient conversion through the MNR.  At the high temperatures at which these resonances occur the overall magnitude of $\mathcal{H}_z$ leading into the level-crossing is larger than it is for the lower-temperature MSW resonance.  Moreover, the Hubble constant, which sets the resonance sweep rate, is larger as well.  At the same time, whereas the off-diagonal weak-interaction potentials are expected to be small leading into either an MSW resonance or an MNR because neutrinos have yet to leave their initial flavor states to any appreciable extent, the off-diagonal vacuum potential is \textit{smaller} at high temperatures.

As one would anticipate based on this argument, the most visible impact of MNR is found when the level-crossing occurs at relatively low temperatures.  In particular, $\eta_{\nu_e} \sim 10^{-6}$ seems to be most clearly affected by the presence of the MNR, which induces non-negligible flavor transformation starting at temperatures near 10 MeV.  Hints of MNR conversion can be seen in our asymmetric-MSW exemplar (Fig.~\ref{2e6intPz}), where it appears that neutrinos are on their way through resonance before abruptly halting their conversion at $T \sim 8$ MeV. Ultimately the overall amount of conversion is limited here too by non-adiabaticity, and the intermingling of the MSW transition with the MNR largely reverses the conversion that does occur.  (Interestingly, for $\eta_{\nu_e} \sim 10^{-6}$ the MNR begets some degree of flavor transformation in the NH as well, in defiance of the general trend for this hierarchy.  The flavor evolution, as it happens, is very similar to that in the IH but with the behavior of neutrinos and antineutrinos exchanged.)

We conclude that conversion through MNR is limited given the parameters adopted in our study.  However, we do not rule out the possibility that a more thorough investigation of the MNR phenomenon in the early universe may reveal sizable effects under appropriate circumstances.  We reiterate that such resonances can exist in the early universe, but that the obstacle to significant transformation is non-adiabaticity.

\subsection{Flavor evolution with quantum damping \label{dampingsec}}

Up to this point the discussion has been couched in the coherent limit.  In reality collisions --- which we model as quantum damping --- will modify these results.  The generic effect of damping is to battle against the development of coherence between the flavors.  It is the combination of oscillations and coherence-erasing damping that leads to depolarization ($P_z$, $\bar{P}_z \rightarrow 0$) and therefore flavor equilibration.  

Indeed, equilibration is generally most effective when flavor transformation is, in the absence of damping, most appreciable.  An immediate consequence is that equilibration is relatively ineffective for most lepton numbers in the NH, which typically fosters only minimal coherent flavor transformation due to the lack of a level-crossing in the MSW region.  While the effects of damping are not entirely insubstantial in the NH, they are usually confined to the relatively placid period during which neutrinos and antineutrinos migrate from their initial flavor eigenstates to the nearby mass eigenstates.

\begin{figure}
\includegraphics[width=.47\textwidth]{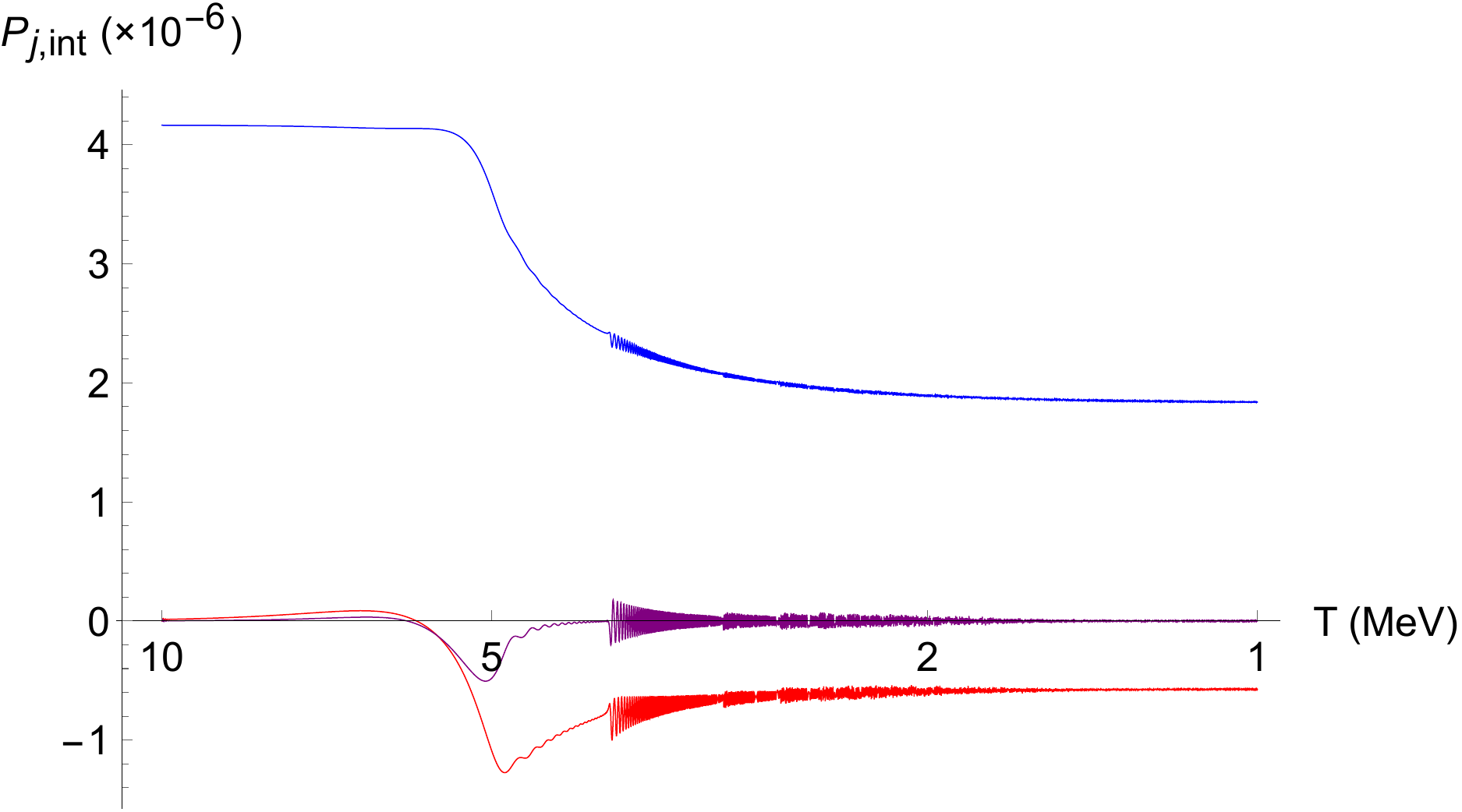}
\caption{(Color online) Minimal transformation (damped): $P_{j, \textrm{int}}$ for $j = z$ (blue, topmost curve at $T = 1$ MeV), $j = x$ (red, bottommost curve at $T = 1$ MeV), and $j = y$ (purple), with the parameters of the minimal-transformation scenario in Fig.~\ref{5e5intPz} (initial degeneracy parameters $\eta_{\nu_e} = 5\times10^{-5}$, $\eta_{\nu_x} = 0$), in the presence of collisional damping.   Antineutrinos undergo qualitatively similar evolution. \label{5e5intPzdamp}}
\end{figure}

Damping is a more potent force in the IH.  The synchronized-oscillation regime, for example, evinces much more efficient depolarization than is witnessed in the NH for the same lepton asymmetries.  At the other end, in the symmetric-MSW regime, depolarization is nearly complete.  But the general trend of efficient depolarization in the IH is not without exception: The development of coherence in the minimal-transformation regime is so limited --- self-coupling is too overpowering for an MSW resonance to occur but too weak to elicit large-amplitude synchronized oscillations --- that damping leaves intact a sizable fraction of the initial asymmetry between the flavors (Fig.~\ref{5e5intPzdamp}).  Previous authors have noted that MSW transitions and synchronized oscillations are vehicles for flavor equilibration, but the existence of a region where neither phenomenon is very compelling, and therefore damping is relatively muted, has not been pointed out before.

In comparing Fig.~\ref{5e5intPzdamp} (damped) with Fig.~\ref{5e5intPz} (coherent), it may come as a surprise that equilibration is not \textit{less} substantial in the damped case than what is shown in Fig.~\ref{5e5intPzdamp}.  The reason is that the flavor evolution plays out in a hierarchy of scales in which the oscillation length is smaller than the mean free path, which in turn is smaller than the MSW resonance width that would obtain for $\mathcal{H}_\nu = 0$.  The picture is this: In the MSW region neutrinos and antineutrinos partially convert flavor, much as they do in the absence of damping.  The polarization vectors accordingly swing away from the flavor axis, and as they do so damping shrinks $\vec{P}_T$.  But due to the high oscillation frequency relative to the scattering rate, the change in $| \vec{P}_T |$ is quickly redistributed over all of the components of $\vec{P}$, so that rather than being flattened against the flavor axis, the polarization vectors are able to evolve in a manner reminiscent of the coherent case, albeit with shrinking magnitude.  In spite of collisions the modes remain largely synchronized, so that $\vec{P} \sim \vec{\mathcal{H}}_\nu \sim \vec{\mathcal{H}}$ (where $\sim$ indicates that the vectors are roughly parallel) as long as self-coupling dominates.  At low temperatures $\vec{\mathcal{H}}_\textrm{vac}$ takes over and all modes move adiabatically into the upper mass eigenstate, just as they do in the coherent limit.  As a matter of fact, the evolution of $P_{z, \textrm{int}} / | \vec{P} |$ is very similar for the damped and coherent cases; the visual differences between Figs.~\ref{5e5intPzdamp} and \ref{5e5intPz} are primarily a result of the long timescale over which the MSW region extends.

\begin{figure}
\includegraphics[width=.47\textwidth]{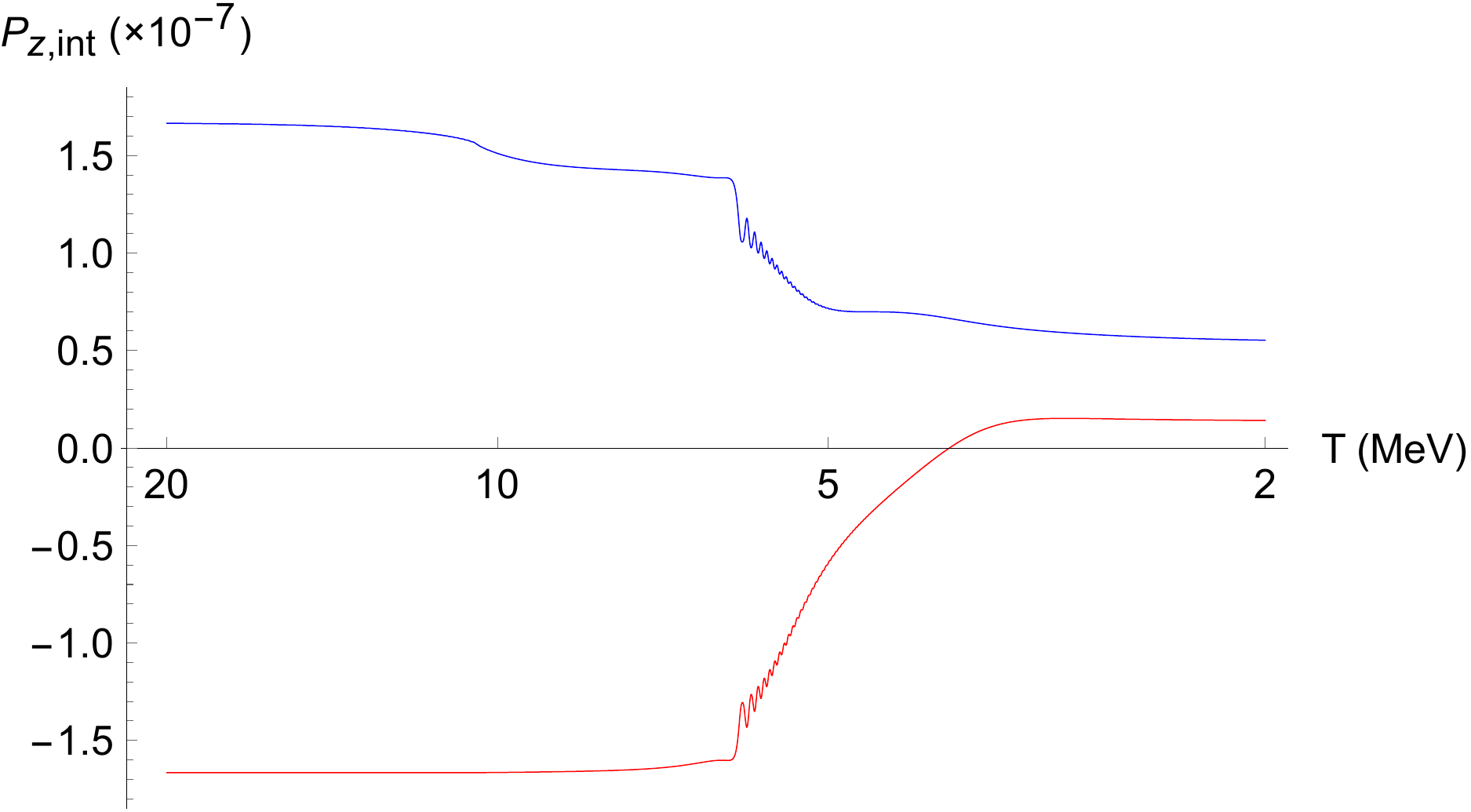}
\caption{(Color online) Asymmetric MSW (damped): $P_{z, \textrm{int}}$ for neutrinos (blue, upper curve at $T = 20$ MeV) and $\bar{P}_{z, \textrm{int}}$ (red), with the parameters of the asymmetric-MSW scenario in Fig.~\ref{2e6intPz} (initial degeneracy parameters $\eta_{\nu_e} = 2\times10^{-6}$, $\eta_{\nu_x} = 0$), in the presence of collisional damping. \label{2e6intPzdamp}}
\end{figure}

\begin{figure}
\includegraphics[width=.47\textwidth]{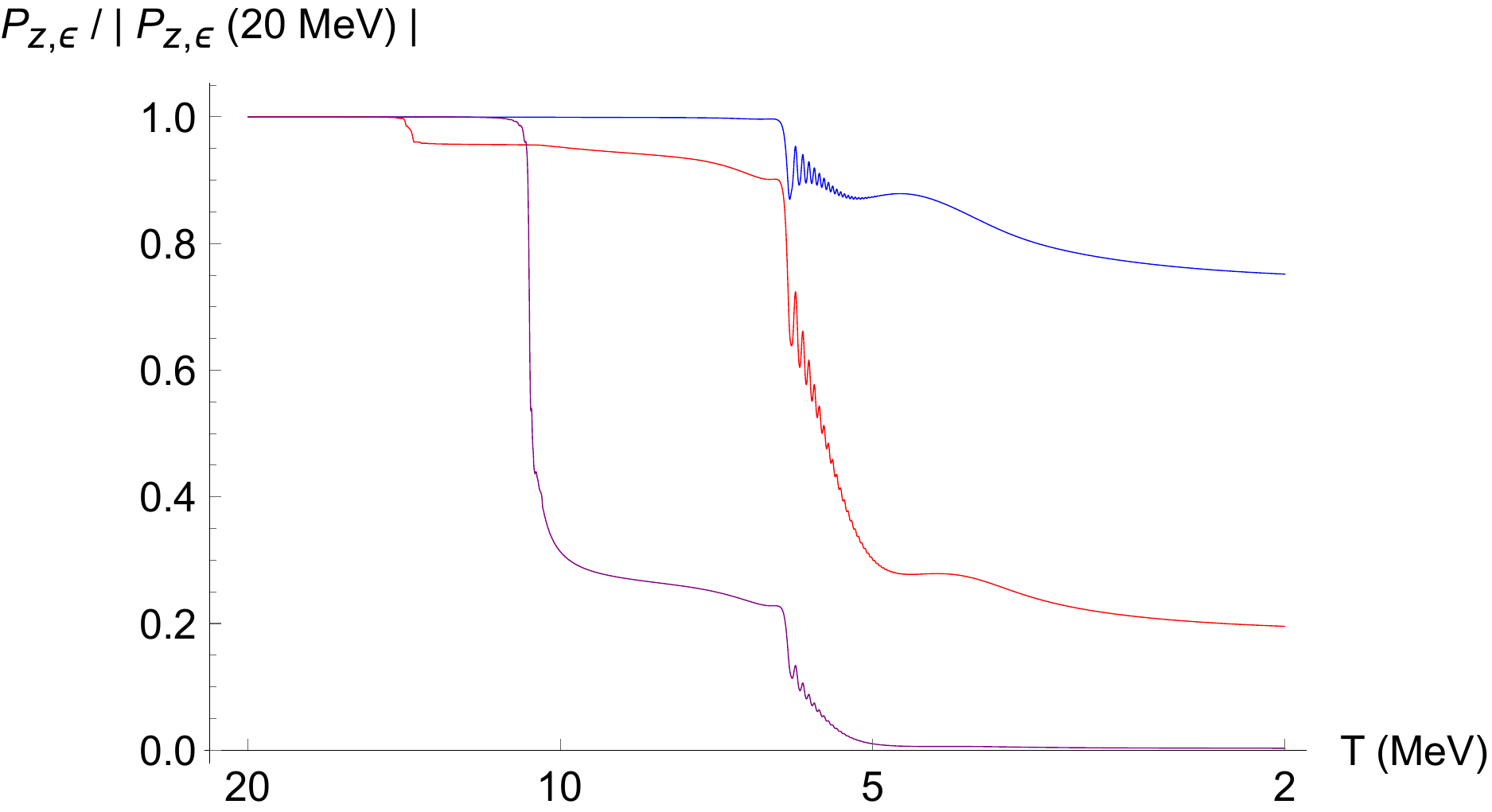}
\caption{(Color online) Asymmetric MSW (damped): $P_{z,\epsilon}$ in the collisionally damped scenario depicted in Fig.~\ref{2e6intPzdamp}, for $\epsilon = 1.15$ (blue, topmost curve at $T = 2$ MeV), $\epsilon = 3.57$ (red), and $\epsilon = 5.99$ (purple, bottommost curve at $T = 2$ MeV).  \label{2e6neudamp}}
\end{figure}

Fig.~\ref{2e6intPzdamp} further illustrates the principle that the degree of depolarization is related to the degree of flavor transformation that takes place in the coherent limit.  For $\eta_{\nu_e} = 2 \times 10^{-6}$, $P_{z,\textrm{int}}$ at weak-decoupling temperatures is $\sim 1/3$ of its initial value at $T \gtrsim 20$ MeV, whereas $\bar{P}_{z,\textrm{int}}$ only retains $\sim 1/12$ of its initial magnitude and manages to change its sign.  The damping of antineutrinos takes place almost entirely during the $\mathcal{H}_\nu$-mediated MSW resonance, of which the small residual $\bar{P}_{z,\textrm{int}}$ is a consequence.  The damping of neutrinos, on the other hand, is more complicated (Fig.~\ref{2e6neudamp}).  Low- and medium-energy modes damp through the MSW region, with greater depolarization associated with greater $\epsilon$, but the high-energy modes undergo damping both through the MSW region and the MNR that occurs at $T \sim 10$ MeV.  Since the scattering rate increases rapidly with temperature, the effectiveness of damping is amplified at the MNR.

Although collisional damping has traditionally been employed in studies of lepton asymmetries, it is nonetheless wanting in realism.  As some authors have noted \cite{pastor2009, mangano2011}, modeling incoherent scattering strictly through this traditional off-diagonal damping term is dubious inasmuch as thermal equilibration requires scattering processes that shuffle neutrinos between energy bins, which such a term cannot provide.  More specifically, if collisions are taken simply to impose damping ($\mathcal{C} \rightarrow -\mathcal{D}\vec{P}_T$ in the polarization-vector language), then one can show that depolarization is inconsistent with the preservation of Fermi--Dirac spectra.

To see that this is so, suppose that at some initial temperature $T_1$ the neutrino gas is in thermal equilibrium with the plasma.  Then, according to our normalization of $\rho$, $P_0$ is the sum of the Fermi--Dirac equilibrium spectra that obtain at this temperature: $P_0 \left( \epsilon, T_1 \right)= f \left( \epsilon, \eta^i_{\nu_e} \right) + f \left( \epsilon, \eta^i_{\nu_x} \right)$.  Suppose also that at some lower temperature $T_2$ damping has achieved complete depolarization: $P_z \left( \epsilon , T_2 \right) \approx 0$ for all $\epsilon$.  Using the fact that coherent evolution and quantum damping both preserve $\textrm{Tr} \rho = P_0$, it follows that
\begin{align}
\rho_{ee} \left( \epsilon , T_2 \right) &= \frac{P_z \left( \epsilon , T_2 \right) + P_0 \left( \epsilon , T_2 \right)}{2} \notag \\
&= \frac{f \left( \epsilon , \eta^i_{\nu_e} \right) + f \left( \epsilon , \eta^i_{\nu_x} \right)}{2}. \label{dampdistort}
\end{align}
Since the average of two Fermi--Dirac spectra is not in general another Fermi--Dirac spectrum, this result implies that the $\nu_e$ distribution function $\rho_{ee}$ picks up distortions from Fermi--Dirac as the polarization vectors shrink to zero --- a troubling conclusion if $P_z$ goes to 0 at high enough temperature that neutrinos must still be in thermal equilibrium.

The consequences of Eq.~\eqref{dampdistort} are borne out numerically: Because the damping term is proportional to neutrino momentum, it engenders a spectral feature wherein higher-energy modes undergo greater depolarization than their lower-energy counterparts.  This feature would be smoothed out somewhat by a more rigorous treatment of incoherent scattering, but it is also indicative of the non-trivial evolution of a system of neutrinos toward equilibrium. Spectral distortions associated with the flavor-equilibration process may compound those known to be generated thermally through the overlapping epochs of $e^\pm$ annihilation and weak decoupling.

The crucial missing ingredient that enforces thermal equilibrium is momentum-changing scattering, which is disallowed when collisions are modeled strictly as quantum damping.  In this vein, the need for a detailed treatment of incoherent scattering was emphasized by Wong \cite{wong2002}, who cautioned that the extent of flavor equilibration depends on how collisions are implemented.  To date, the most sophisticated analyses of flavor evolution with a lepton asymmetry are those performed by the authors of Refs.~\cite{pastor2009, mangano2011, mangano2012, castorina2012}, who have combined an off-diagonal damping term with classical Boltzmann collision integrals along the diagonals of $\mathcal{C}$.  By revealing a wider range of possible coherent phenomena than has hitherto been recognized, our results buttress the need for continued progress in this direction.

As the findings of Ref.~\cite{grohs2016} have demonstrated, BBN calculations that self-consistently couple neutrino transport to the thermodynamics of the plasma yield changes in the predicted primordial abundance of D --- relative to the case of instantaneous neutrino decoupling --- that are an order of magnitude larger than they would be if the non-linear feedback between the neutrinos, plasma, and nuclides were omitted.  The calculations of Ref.~\cite{grohs2016}, however, were performed with zero lepton number in the classical Boltzmann limit.  Ref.~\cite{desalas2016}, meanwhile, tackled the full problem of oscillations and quantum collision integrals but was predicated on the assumption of zero lepton asymmetry.  A similar approach to the complete quantum kinetic equations \cite{volpe2013, zhang2013, vlasenko2014, serreau2014, kartavtsev2015}, including fully realistic quantum collision integrals \cite{blaschke2016} \textit{and} a nonzero lepton asymmetry, may divulge signatures of flavor evolution in the early universe that are currently believed to be unobservable.

\section{Conclusion \label{conclusion}}

In this paper we have numerically solved the coherent equations of motion governing neutrino flavor transformation in the early universe with a range of initial lepton asymmetries.  In so doing we have discovered that beneath the current constraints on the lepton number there lurks a menagerie of possible coherent flavor phenomena, which we have sectioned off into five distinct regimes.  Starting from a lepton asymmetry comparable to the present bound and moving down to the realm of negligible self-coupling, these regimes are as follows: (1) Large synchronized oscillations, (2) minimal transformation, (3) asymmetric MSW, (4) partial MSW, and (5) symmetric MSW.  The existence of these regimes is a testament to the richness of the nonlinear problem of flavor evolution in a dense, expanding environment.  And as we have demonstrated, this richness is not entirely erased by collisional damping --- a finding that points to the merits of further study of this problem with quantum kinetics that go beyond the approximations employed here.

To explain the phenomena observed in our numerical results we have employed the conceptual apparatus of (non-)adiabatic level-crossings and the well-established understanding of synchronized evolution as a collective mode that emerges when the self-coupling potential is dominant.  Yet we also contend that in fact the minimal-transformation regime, which occurs for lepton asymmetries on the order of $\sim 5 \times 10^{-5}$, points to the limitations of these concepts.  The distinctive absence of flavor conversion in this regime is due to it encompassing lepton asymmetries that are strong enough to eliminate level-crossings in the MSW region but not strong enough for $\mathcal{H}_\nu$ to develop the dominant off-diagonal components needed for large-amplitude synchronized oscillations, much less for $\mathcal{H}_\nu$ to bind the individual modes sufficiently for a synchronized MSW transition to take place.  As far as we are aware, a convincing analytical understanding of this regime does not currently exist.  We note again that it is an intriguing coincidence that the range of lepton numbers most consistent with an interpretation of the unidentified X-ray line reported in Refs.~\cite{bulbul2014, boyarsky2014} falls within this regime, which is the one most resistant to damping-induced flavor equilibration.

We have also reported for the first time the existence of an MNR in the early universe.  The influence of the resonance on coherent flavor evolution is very modest except for a small range of lepton asymmetries for which the level-crossing occurs shortly before the MSW region.  Its presence is accentuated by damping, which capitalizes on the coherence developed through the resonance.  We have found that adiabaticity restricts the amount of flavor conversion through the MNR, but mixing through the $\delta m^2_\odot$ channel, which has MSW resonances at lower temperatures than those studied here, may permit more adiabatic circumstances.

The sub-constraint lepton asymmetries we have investigated are, by definition, thought to lie presently out of reach of observation.   Nonetheless, the diversity of flavor phenomena revealed in this study may have unrecognized implications for BBN.  The current era is one of precision cosmology, with 30-meter-class telescopes \cite{spyromilio2008, sanders2013, bernstein2014}, forthcoming spectroscopic galaxy surveys \cite{font2014, abazajian2015, kim2015}, and a Stage-IV CMB experiment \cite{abazajian2015, wu2014} at the vanguard --- to name just a few.  Impressive advances in determinations of $N_\textrm{eff}$, $Y_P$, $\left[ \textrm{D} / \textrm{H} \right]$, and other cosmological observables are on the horizon.  These measurements promise to provide new insights, but exploiting them thoroughly will require a scrupulous treatment of neutrino evolution.  It remains to be seen whether a solution of the full quantum kinetic equations coupled to BBN will unearth traces of the physics presented in this study.

\begin{acknowledgments}
We thank Brad Keister, Chad Kishimoto, and Amol Patwardhan for helpful conversations.  This work was supported by NSF Grant No. PHY-1307372 at UC San Diego; by the Los Alamos National Laboratory Institute for Geophysics, Space Sciences and Signatures Subcontract No. 257842; by the Los Alamos National Laboratory Institutional Computing Program, under U.S. Department of Energy National Nuclear Security Administration Contract No. DE-AC52-06NA25396; by the Los Alamos National Laboratory LDRD Program; and by the U.S. Department of Energy Office of Science Graduate Student Research (SCGSR) Program.
\end{acknowledgments}

\bibliography{lepton}

\end{document}